\documentclass[12pt,draftclsnofoot,onecolumn]{IEEEtran}
\IEEEoverridecommandlockouts

\usepackage{graphicx}
\usepackage{subfigure}
\usepackage{amsmath,amsthm,amsfonts,amssymb}
\usepackage{cite}
\usepackage{bm}
\usepackage{url}
\usepackage{array}
\usepackage{color}
\usepackage{booktabs}
\usepackage{multirow}
\usepackage{enumerate}

\usepackage[noend]{algpseudocode}
\usepackage{algorithmicx,algorithm}

\allowdisplaybreaks[4]

\theoremstyle{plain}
\newtheorem{thm}{Theorem}
\newtheorem{lem}[thm]{Lemma}

\newtheorem{prop}[thm]{Proposition}

\newtheorem*{rem}{Remark}

\newenvironment{NewProof}{{\noindent\it Proof.}}{\hfill $\blacksquare$\par}
\providecommand{\myceil}[1]{\left \lceil #1 \right \rceil }
\providecommand{\myfloor}[1]{\left \lfloor #1 \right \rfloor }

\begin{document}
\title{Sporadic Ultra-Time-Critical Crowd \\ Messaging in V2X}

\author{
Yulin~Shao,~\IEEEmembership{Student Member,~IEEE},
Soung~Chang~Liew,~\IEEEmembership{Fellow,~IEEE},
Jiaxin~Liang,~\IEEEmembership{Student Member,~IEEE}
\thanks{Y. Shao, S. C. Liew and J. Liang are with the Department of Information Engineering, The Chinese University of Hong Kong, Shatin, Hong Kong (e-mail: \{sy016, soung, lj015\}@ie.cuhk.edu.hk). This paper was presented in part in ICC 2018~\cite{confVersion}.}
}

\maketitle

\begin{abstract}
Life-critical warning message, abbreviated as {\emph{warning}} message, is a special event-driven message that carries emergency information in Vehicle-to-Everything (V2X).
Three important characteristics that distinguish warning messages from ordinary vehicular messages are {\emph{sporadicity}}, {\emph{crowding}}, and {\emph{ultra-time-criticality}}.
Specifically, warning messages come only once in a while in a sporadic manner; however, when they come, they tend to come as a crowd and they need to be delivered in short order.
This paper puts forth a medium-access control (MAC) protocol for warning messages.
The overall MAC protocol operates by means of {\emph{interrupt-and-access}}.
To circumvent potential inefficiency arising from message sporadicity, we adopt an override network architecture whereby warning messages are delivered on the spectrum of the ordinary vehicular messages.
A vehicle with a warning message first sends an interrupt signal to pre-empt the transmission of ordinary messages, so that the warning message can use the wireless spectrum originally allocated to ordinary messages. In this way, no exclusive spectrum resources need to be pre-allocated to the sporadic warning messages. Following the interrupt, for transmissions of ultra-time-critical crowd messages, we employ advanced channel access techniques to ensure reliable message delivery within an ultra-short time in the order of ${10}$ ms.
\end{abstract}


\section{Introduction}
With the explosive growth of vehicles on road, safety has become a major concern for future intelligent transportation systems (ITS) \cite{5GPPP}.
Statistical data show that the number of crashes in the United States is nearly $6$ million each year \cite{NHTSA}. Vehicle-to-Everything, abbreviated as V2X, is a promising means to cut the road toll \cite{v2xbook}. Through V2X, all the entities on the road (e.g., vehicles, road side units and pedestrians) are connected, hence, they can exchange safety messages and cooperate to prevent road accidents or cut down fatality and injury rates when they do occur.

Safety messages in V2X can be classified into two categories \cite{v2xbook}:
1) {\it heartbeat} messages. Each on-road node periodically broadcasts heartbeat messages to declare its existence, current state and environment information. Receiving nodes can then evaluate whether there are hazards from information disseminated by transmitters and data gathered from the environment.
2) {\it event-driven} messages. Safety in V2X is not limited to passive evaluation of the received heartbeats. An on-road node encountering unexpected events could actively broadcast event-driven messages so that the surrounding nodes can respond quickly. Typical events that may induce event-driven messaging include lane change, roadwork, ambulance approach, to name a few.

Among event-driven messages, life-critical warning messages (hereinafter, referred to as {\it warning} messages) deserve particular attention. Warning messages are triggered by extreme traffic emergencies that are likely to cause casualties, e.g., hard braking on the highway, imminent crash, and swerving vehicles at the opposite lane.
Typically, a warning message contains the following data \cite{DSRC}: node ID ($4$ bytes), message generation time ($4$ bytes, modulo one minute, with resolution $1~\mu s$), message type ($2$ bytes, e.g., braking, acceleration, steering) and message attributes ($14$ bytes, e.g., for braking message type, the attributes could contain brake force, current vehicle speed and wheel state) for an aggregate of $24$ bytes.

Three important characteristics that distinguish warning messages from ordinary vehicular messages are as follows:
\begin{enumerate}
\item They are rare and sporadic. Statistics indicate that there are on average $1.04$ fatal crashes every $100$ million miles a vehicle travels \cite{NHTSA}.
\item Warning messages are short but multiple warning messages may arrive as a crowd in a batch. This is because a single emergency event can trigger multiple emergency responses from multiple nearby nodes. As a result, these emergency nodes (typically less than $30$) can broadcast multiple warning messages simultaneously.
\item They must be delivered with high certainty in short order. According to the automotive white paper from 5G-PPP \cite{5GPPP}, the maximum tolerable end-to-end delay of these safety-of-life messages is $10$ ms, and the maximum tolerable message loss rate within $10$ ms is $10^{-4}$.
\end{enumerate}
We refer to these three message characteristics as {\it sporadicity}, {\it crowding}, and {\it ultra-time-criticality}.

\begin{figure}[t]
  \centering
  \includegraphics[width=0.75\columnwidth]{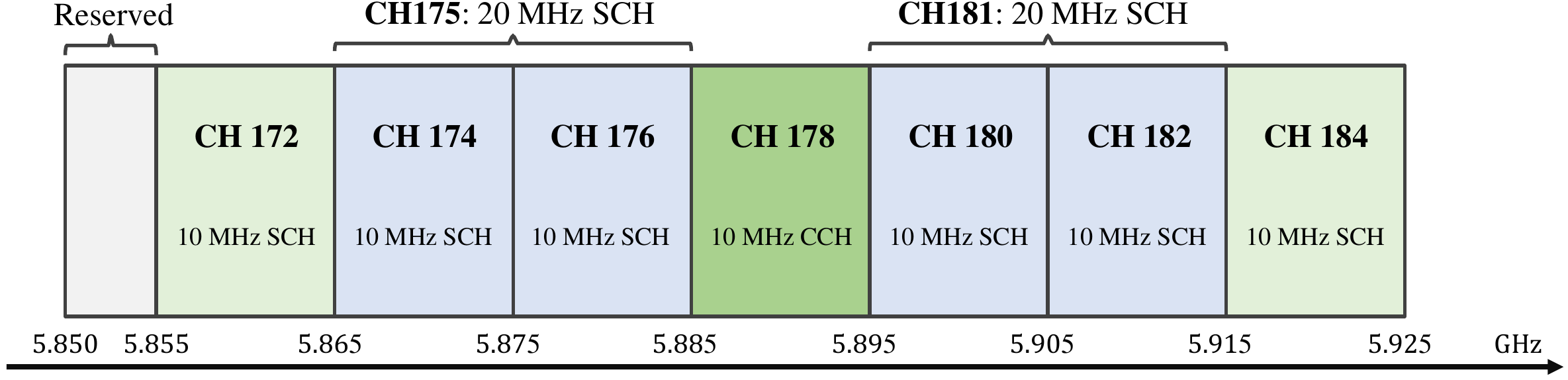}\\
  \caption{The $75$ MHz ``5.9 GHz band''. Channel $178$ is a control channel dedicated for safety-message transmission. The other six channels are service channels, in which channel 172 and 184 are reserved for future advanced applications. Service channels can be used to transmit both safety and non-safety messages.}
  \setlength{\belowcaptionskip}{-10pt}
  \label{Fig1}
\end{figure}

In V2X, safety messages are disseminated by simple means of one-hop broadcast. Multiple access control (MAC) designs are especially crucial if the stringent delay and reliability requirements are to be met. As shown in Fig. \ref{Fig1}, the Federal Communications Commission (FCC) in the United States allocates $75$ MHz ``5.9 GHz band'' for V2X communication \cite{DSRC,CITS}, based on which many MAC protocols have been proposed and developed to support safety-message broadcasting \cite{DSRC,CITS,LTEforV2X,ProSe,VeMAC,NEWShort,NEWLong,LTE5GV2X}. However, existing schemes are designed primarily for heartbeat messages and conventional event-driven messages. When it comes to sporadic ultra-time-critical crowd messaging, none of them can meet the stringent delay and reliability requirements (see section \ref{sec:II} for more details).

To fill this gap, this paper puts forth an interrupt-and-access MAC protocol tailored for the delivery of life-critical warning messages.
Two underpinnings of our MAC protocol are as follows:
\begin{enumerate}
\item \textbf{Interrupt}: To prevent inefficient spectrum usage due to sporadicity of warning messages, we do not dedicate exclusive spectrum to them. Rather, we build our MAC protocol upon an override network architecture whereby wireless spectrum originally allocated to regular vehicular network is used to deliver warning messages only when they appear.
    A vehicle with warning messages will first send an interrupt message to pre-empt the transmission of regular vehicular messages so that the warning message can follow after that.
\item \textbf{Access}: To address crowding and ultra-time-criticality, we use advanced channel access techniques to ensure reliable delivery of crowd warning messages within the stringent delay target.
    In a life-threatening situation, multiple vehicles may have life-critical messages to send.
    A channel access protocol that does not incur excessive hand-shaking overhead to coordinate the transmissions of these vehicles on the shared spectrum is critical if the stringent delay target is to be met.
\end{enumerate}

For wireless interrupt, we devise an interrupt mechanism that makes use of spread spectrum sequences \cite{AlphaSeq} on $5.8$ GHz Industrial Scientific Medical (ISM) band as interrupt signals.
Given a tolerable false alarm rate (FAR), the optimal Neyman-Pearson detector is devised to minimize the misdetection rate (MDR) for the detection of interrupt signals.
Numerical and simulation results show that the misdetection rate (MDR) of the interrupt signals can be very small provided that the interrupt sequences are long enough, e.g., when the signal to interference plus noise ratio (SINR) is $-28.2$ dB, a $0.43$ ms sequence ($64512$ symbols, $150$ MHz) can guarantee an MDR of $10^{-6}$.

For channel access, we investigate two uncoordinated channel access schemes for reliable multiple access.
Targeting for a $\bm{10^{-4}}$ message loss rate in our set-up, a simple multi-replica ALOHA scheme can support up to $11$ nodes.
If the number of transmitters exceeds $11$, a more advanced (and more complex) coded ALOHA scheme can potentially support up to $120$ nodes while keeping the message loss rate lower than $10^{-4}$.


\section{State-of-The-Art V2X MAC Protocols}\label{sec:II}
Existing MAC protocols for V2X communications operate in either a distributed or a centralized manner \cite{LTEforV2X,ProSe}.
Centralized MAC designs have certain limitations:
1) Infrastructure could be a single point of failure. These MAC protocols may not function when infrastructure failure occurs or when vehicles are out of the coverage of the infrastructure (e.g., in blind zone, tunnels, and underground parking lots).
2) The coordination-based framework, e.g., schedule-before-transmit \cite{LTEforV2X}, does not fit delay-sensitive applications, owing to the extra delay and overhead consumed.
Distributed self-organizing MAC designs are in general more suitable for ultra-delay-sensitive warning messages \cite{5GPPP}.

\subsection{IEEE 802.11p and IEEE 802.11bd}
Dedicated short-range communication (DSRC) \cite{DSRC} refers to the sets of standards on the 5.9 GHz band. The MAC protocol in DSRC, i.e., IEEE 802.11p \cite{802.11p}, is an amendment from IEEE 802.11a with enhanced distributed channel access (EDCA) Quality-of-Service (QoS) extension. In 802.11p, both heartbeat and event-driven messages share the $10$ MHz control channel (CCH) by means of carrier sensing multiple access (CSMA). In particular, different types of messages are assigned with different priorities: high-priority messages have smaller interframe spacing and backoff waiting time so that they have priority over low-priority messages in channel access.

IEEE 802.11bd \cite{NEWLong} is an ongoing evolution of the IEEE 802.11p. The main design objective of 802.11bd is to support advanced V2X applications such as vehicle platooning, advanced driving, extended sensors, and remote driving. Most of the improvements of 802.11bd over 802.11p are in the physical layer, e.g., higher-order modulation, reduced subcarrier spacing, midamble for accurate channel estimation, low-density parity-check (LDPC) codes, and space-time block coding (STBC). At the MAC layer, however, 802.11bd reuses all the contention parameters for different EDCA categories for backward compatibility with 802.11p devices \cite{NEWLong,NEWShort}.

There are three main reasons why 802.11p and 802.11bd are not suitable for sporadic ultra-time-critical crowd messaging, even if we assign the highest priority to warning messages.
\begin{enumerate}
\item Delay concern -- In both protocols, messages with different priorities share the same control channel. When high-priority warning messages are generated, a low-priority message may be in the midst of occupying the channel. As a result, the warning messages must wait until the channel is idle before transmission. Furthermore, even if the channel is idle, multiple warning messages with the same high priority may compete for the channel simultaneously, leading to a high collision rate that may significantly increase the delay.
\item Lack of acknowledgment (ACK) -- Warning messages are broadcasted for all vehicles in the vicinity of the warning-message generating vehicle. However, requiring an ACK from each one-hop neighbor of the broadcaster (potentially hundreds of nodes) can be highly costly. Thus, 802.11p and 802.11bd do not have ACK for broadcast messages, and nodes cannot detect collisions \cite{NEWLong}. As a result, there is no retransmission in 802.11p, hence the packet transmission is highly unreliable. In 802.11bd, nodes can retransmit for at most 3 times. In particular, retransmissions can be sent a) following the initial transmission, or b) by separate contention processes. However, scheme a) triples (or doubles, depending on the number of retransmissions) the duration of a channel occupancy, thereby reducing the throughput. Also, a single collision of the initial transmission means all its retransmissions collide. On the other hand, scheme b) increases the collision probability since the number of packets of the same priority is now tripled.
\item Hidden node problem -- To tackle the hidden node problem, RTS/CTS handshaking is implemented in conjunction with CSMA in IEEE 802.11a. However, in 802.11p and 802.11bd, the hidden node problem is left unsolved \cite{NEWLong}, because for broadcast messages, the frequent RTS/CTS handshakes consume too many resources. As a result, a warning broadcast message may collide with another warning broadcast message two hops away, leading to packet loss.
\end{enumerate}

\subsection{TDMA-based MAC}
Vehicular MAC protocols may also be based on time division multiple access (TDMA). Two representative examples are ADHOC MAC \cite{ADHOCMAC} and its multi-channel evolution VeMAC \cite{VeMAC}.

As with IEEE 802.11p, VeMAC use the control channel (CH 178) in the 5.9 GHz band for both heartbeat and event-driven messages. This $10$ MHz channel is assumed to be time-slotted, and every $M$ slots are grouped together as a frame.
In VeMAC, each node occupies at least one slot in every frame for the broadcast of its heartbeat message. If a node has an event-driven message to broadcast, it will need to acquire one more slot. In particular, the slots a node occupies must be different from the slots occupied by any of its neighbors within two hops (this guarantees that there is no hidden node problem). How nodes within two hops coordinate with each other and occupy different slots is an essence of VeMAC.

To enable sporadic ultra-time-critical crowd messaging in the context of VeMAC, all nodes can reserve one slot every $10$ ms to cater for the rare occasion when they have warning messages to broadcast. However, simple calculation indicates that this is not viable. Assuming the slot duration is $50~\mu s$, and therefore, there are $200$ slots available in $10$ ms, if there are $200$ nodes within two hops (in practice could be up to $1000$), then all the slots are reserved by these nodes for warning messaging alone. Even if we assumed sparse nodes, reserving resources for warning messaging is quite inefficient, because warning messages are rare and sporadic.

Instead of exclusive reservation of slots, a node could attempt to acquire a slot only upon the generation of a warning message. However, slot acquisition under VeMAC takes one or more frames (a frame usually lasts for $100$ ms \cite{VeMAC}), because the transmitter must wait for all its one-hop neighbors' ACKs to make sure the new slot is free for it to use. Worse still, when there are $K$ nodes with warning messages, the interaction process for them to acquire $K$ different new slots can take an inordinate amount of time.

\subsection{C-V2X and NR-V2X}
Another evolution path of V2X communications is standardized by the Third Generation Partnership Project (3GPP) \cite{LTE5GV2X}. The main objective is to leverage 4G long term evolution (LTE) and 5G new radio (NR) connectivity to enable V2X applications. The premiere version, defined in Releases 14 and 15, is named Cellular V2X (C-V2X) \cite{LTE5GV2X,NEWShort}, and the latest version underway, to be completed in Releases 16, is named New Radio V2X (NR-V2X) \cite{NEWLong}. Both C-V2X and NR-V2X support two operation modes, one is centralized communications coordinated by the base stations (i.e., eNodeB in LTE and gNodeB in NR) in the licensed spectrum, and the other is direct (sidelink) communications between vehicles in the 5.9 GHz band\footnote{To be specific, direct sidelink vehicular communications in the out-of-coverage scenario are defined in C-V2X sidelink mode 4 and NR-V2X sidelink mode 2, respectively.}. The second operation mode is more suitable for ultra-delay-sensitive warning messages.

The MAC protocol of C-V2X is tailored for periodic basic safety messages. Specifically, each vehicle senses and analyzes the channel occupation during a sensing window of one second. Whenever a vehicle has a message to send, it will randomly select the physical resource blocks that are not busy based on its cognitions in the past one second. An underlying assumption of this sensing-based design is that the future can be inferred from the past. This assumption, however, does not apply to our problem since warning messages arrive sporadically and aperiodically.

To support the reliable delivery of aperiodic messages, the NR-V2X study item was launched in 5G Release 16 (by July 2020, Rel-16 Stage 3 freeze was approved) \cite{NEWShort}. The primary MAC improvements of NR-V2X over C-V2X are as follows. 1) In addition to broadcast, unicast and groupcast were introduced; HARQ was introduced for unicast and groupcast such that retransmission is possible. 2) Aperiodic messages can use a short-term sensing window (e.g., listen before talk as in Wi-Fi) instead of a fixed one-second sensing window. 3) An idea of preemptive resource scheduling is proposed for critical messages. Specifically, vehicles with a critical message can send a preemptive indicator signal to other vehicles so that vehicles with reserved resources for less critical messages can release them to support the quick transmission of critical messages.

The third improvement above in recent developments of NR-V2X is relevant to our study in this paper. The idea of preemptive transmission coincides with our idea of interrupt-and-access mechanism tailored for life-critical warning messages. While no design instructions for pre-emption was defined in Rel-16, our design in this paper is the first realization of the preemptive transmission, to the best of our knowledge, and can potentially be incorporated into NR-V2X as a solution to enable sporadic ultra-time-critical crowd messaging.

\section{An Override Architecture}\label{sec:III}
\begin{figure}[t]
  \centering
  \includegraphics[width=0.7\columnwidth]{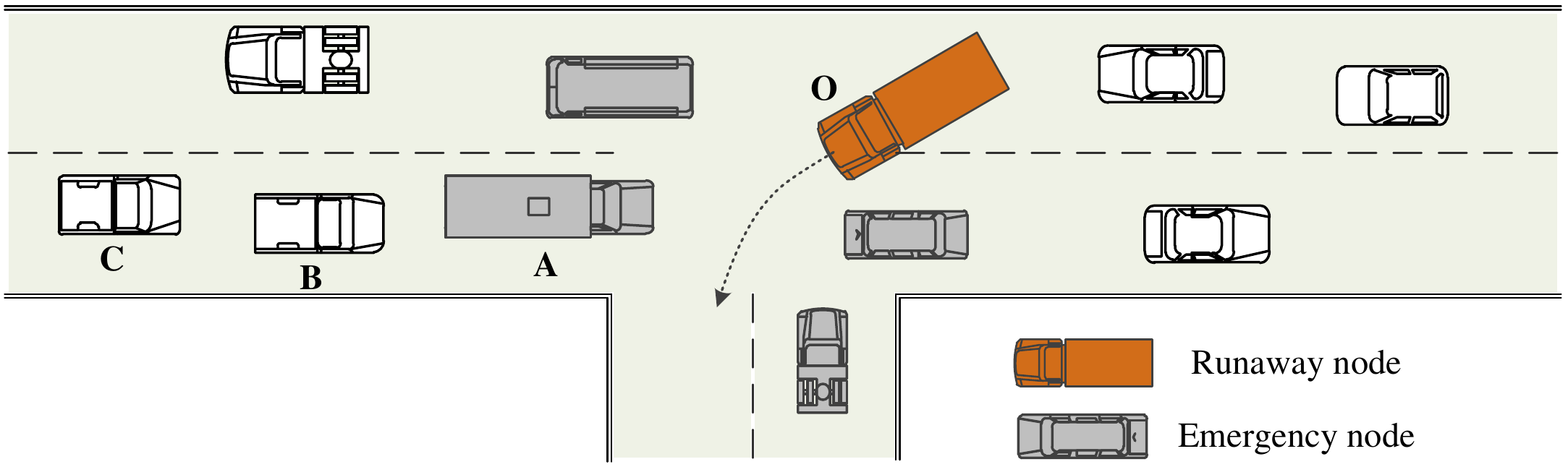}\\
  \caption{The runaway vehicle $O$ violates the traffic light. This accident triggers the urgent reactions of $K$ nodes ($K=4$ in this figure), and each of them generates a warning message to warn its nearby nodes.}
\label{Fig2}
\end{figure}

Let us consider a typical V2X scenario where $K_\textup{all}$ on-road nodes are communicating with each other on the 5.9 GHz band in an ad-hoc manner. Each node is equipped with two sets of half-duplex transceivers TRX$_1$ and TRX$_2$. TRX$_1$ is aligned to the $10$ MHz control channel (CH 178), on which nodes exchange heartbeat or conventional event-driven messages to get an overall perception of the environment. TRX$_2$ is aligned to the $40$ MHz service channels (CH 175 and 181), on which nodes exchange non-safety messages, e.g., infotainment messages.

As illustrated in Fig. \ref{Fig2}, an accident suddenly happens: a runaway vehicle $O$ violates the traffic light and runs to an opposite lane. This accident triggers the urgent reactions of $K$ nearby nodes, and each of them generates a life-critical warning message to warn its nearby nodes. For example, node $A$ brakes hard, triggering a warning message informing its neighbors (e.g., nodes $B$ and $C$) of its emergency braking caused by the runaway node $O$, so that they could react in time to avoid further crashes.

These life-critical warning messages have stringent delay and reliability requirements. In this sense, we may need to assign them sufficient time-frequency resources, so that the stringent QoS requirements can be met.
On the other hand, warning messages are sporadic and arrive once in a long while, hence, assigning them exclusive resources is highly inefficient, because these resources are wasted most of the time in the absence of life-critical events. This motivates us to build the warning-message MAC protocol upon an override network architecture, where life-critical warning messages share the $40$ MHz service channels with non-safety messages. In non-emergency situations, non-safety messages are the primary users on the service channels. {\it When an emergency arises, high-priority warning messages will override non-safety messages and seize the service channels}\footnote{In practice, warning messages can override the whole $40$ MHz bandwidth or part of the bandwidth of the service channels, e.g., override only the $20$ MHz channel 181, so that non-safety messages would not be totally deprived of services.}.

\begin{figure}[t]
  \centering
  \includegraphics[width=0.6\columnwidth]{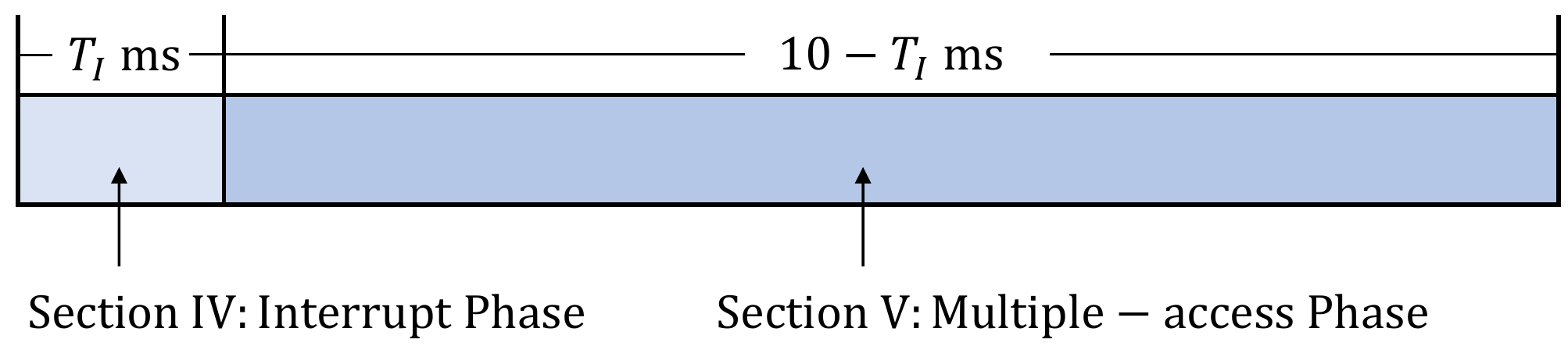}\\
  \caption{The override architecture operates by means of interrupt-and-access. The time consumed by the two phases is limited to $10$ ms.}
\label{Froadmap}
\end{figure}

As shown in Fig.~\ref{Froadmap}, our override architecture operates by means of interrupt-and-access: nodes with incoming warning messages send interrupt signals to nodes transmitting non-safety messages to pre-empt them so that the nodes with warning messages can broadcast on the service channels. Sections IV and V provide the details of wireless interrupt and channel access, respectively.

\section{Wireless Interrupt}\label{sec:IV}
Interrupt is a technique widely used in computer systems for multitasking with different priorities.
Specifically, an incoming high-priority task triggers an interrupt signal to the central processor, so that the processor can suspend the currently ongoing low-priority task and process the high-priority task immediately.
Interrupt is rarely used in conventional wireless communication systems. It is, however, useful for our application scenario.

\subsection{Interrupt protocol}
In the vehicular network, if a node wants to broadcast a warning message successfully on the service channels, a prerequisite is that all its one-hop and two-hop neighbors are silent on these channels.
The objective of wireless interrupt in V2X is to silence these neighbors (i.e., neighbors within two hops) in case they are in the midst of transmitting on the service channels. To this end, we define two interrupt signals, a primary interrupt signal (PIS) and a secondary interrupt signal (SIS). The wireless interrupt protocol are given in Algorithm~\ref{InterruptProtocol}.

\begin{algorithm}[t]
\caption{Wireless Interrupt protocol for warning messaging in V2X.}
\label{InterruptProtocol}
\begin{algorithmic}[1]
\State A node with a warning message to transmit first broadcasts a PIS.
\State Any node detecting the PIS then broadcasts a SIS.
\State Any node detecting a PIS or a SIS (except for nodes in step iv below) keeps silent on the service channels for $10$ ms, including nodes that are halfway transmitting on a service channel when the PIS or SIS is detected.
\State Any node that issues PIS, regardless of whether it receives a PIS/SIS from another node, then transmits its warning message during the channel access period following the interrupt period.
\end{algorithmic}
\end{algorithm}

\begin{figure}[t]
  \centering
  \includegraphics[width=0.7\columnwidth]{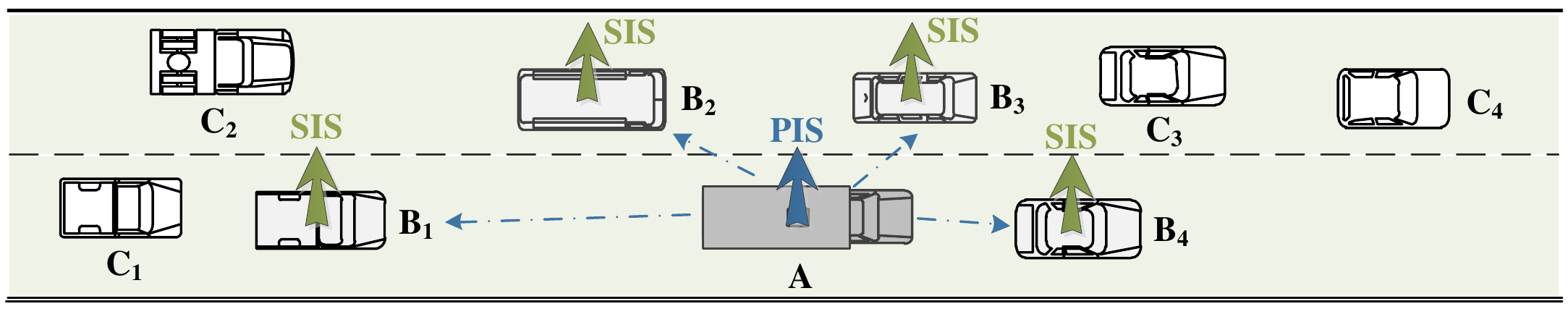}\\
  \caption{Interrupt in V2X. The emergency node $A$ broadcasts the PIS to its one-hop neighbors, and detecting a PIS will trigger the broadcasting of SIS. All node $A$'s neighbors within two hops will keep silent after they receive a PIS or SIS.}
\label{Fig3}
\end{figure}

Let us use the example in Fig.~\ref{Fig3} to expound how the interrupt proceeds. In Fig.~\ref{Fig3}, an emergency node $A$ has a warning message to broadcast. The one-hop neighbors of node $A$ are $\bm{B}=\{B_1,B_2,B_3,B_4\}$, and the two-hop neighbors of $A$ are $\bm{C}=\{C_1,C_2,C_3,C_4\}$.
Following Algorithm~\ref{InterruptProtocol}, node $A$ first broadcasts a PIS to pre-empt other nodes from using the service channels.
Upon detecting the PIS, node $A$'s one-hop neighbors $\bm{B}$ keep silent on the service channels for $10$ ms (step 3). Moreover, the detection of PIS triggers each of them to broadcast an SIS (step 2), so that $A$'s two-hop neighbors $\bm{C}$ can detect the SIS and keep silent on the service channels as well.
After interruption, node $A$'s one-hop and two-hop neighbors are silent on the service channels in the next $10$ ms, and node $A$ can broadcast the warning message in the channel access period safely.
In particular, we specify that
1) any emergency node that issues a PIS will transmit its warning message on the service channels during the channel access period, regardless of whether it receives a PIS/SIS or not.
2) For a non-emergency node that is within two hops of an emergency node, interrupt is successful as long as at least one interrupt signal is detected, whether it is PIS or SIS.

When there are more than one emergency nodes, multiple overlapped PISs and SISs can be broadcasted simultaneously. In this case, a node will detect the multiple PISs and SISs separately, and respond to each of the interrupt signal following the third step in Algorithm~\ref{InterruptProtocol}.

\subsection{Design of interrupt signals}
Given the interrupt protocol, the next problem is how to design the PIS and SIS.
Potentially, there are two realizations of wireless interrupt, one is an in-band realization and the other is an out-of-band realization.

We could interrupt in-band by exploiting special features of non-safety signal on the service channels. For instance, assuming the non-safety messages are carried by OFDM signals, we could transmit the interrupt signal on the guard-band subcarriers\footnote{Another alternative to realize in-band interrupt is full duplex communication. However, full duplex communication requires dedicated full duplex transceivers, i.e., tailor-made RF chips with self-interference cancellation. Interrupt via full-duplex techniques is overkill, because unlike the receiver of a full-duplex link, the receiver of an interrupt does not need to receive a data stream in the reverse direction; it only needs to be able to detect the presence of an interrupt signal.}.

This paper considers out-of-band interrupt. Speci\-fically, we transmit interrupt signals on the $5.8$ GHz ISM band. PIS and SIS are designed as spread-spectrum sequences on this $150$ MHz band, so that they can be detected in the presence of interference.

\subsubsection{Interference on the 5.8 GHz band}
The $5.8$ GHz ISM band ($5.725$-$5.875$ GHz) is a free radio band centered on $5.8$ GHz, the channel characteristics of which is similar to that of the $5.9$ GHz vehicular band. The primary traffic on the 5.8 GHz band is Wi-Fi signal, and Wi-Fi are commonly deployed indoors.

To evaluate the interference of indoor Wi-Fi signal to our outdoor interrupt signal, we conducted an experiment over our campus to capture $5.8$ GHz Wi-Fi signal using USRP X310 (with BasicTX daughter board). The experimental data indicates that in the outdoor environment, 1) most of the time, nothing can be detected on the $5.8$ GHz band; 2) when Wi-Fi signal was detected on the $5.8$ GHz band, the signal power was much lower than the indoor power.

In one experiment, an access point (AP, Linksys EA6900) was deployed indoors. The AP used Wi-Fi channel 153 (5.765 GHz) with $20$ MHz channel bandwidth. We measured the received Wi-Fi signal intensities from the AP at two locations. The first location was indoor ($5$ meters from the AP, LOS), and the second location was outdoor (straight distance $20$ meters from the AP, NLOS).

\begin{figure}[t]
  \centering
  \includegraphics[width=0.7\columnwidth]{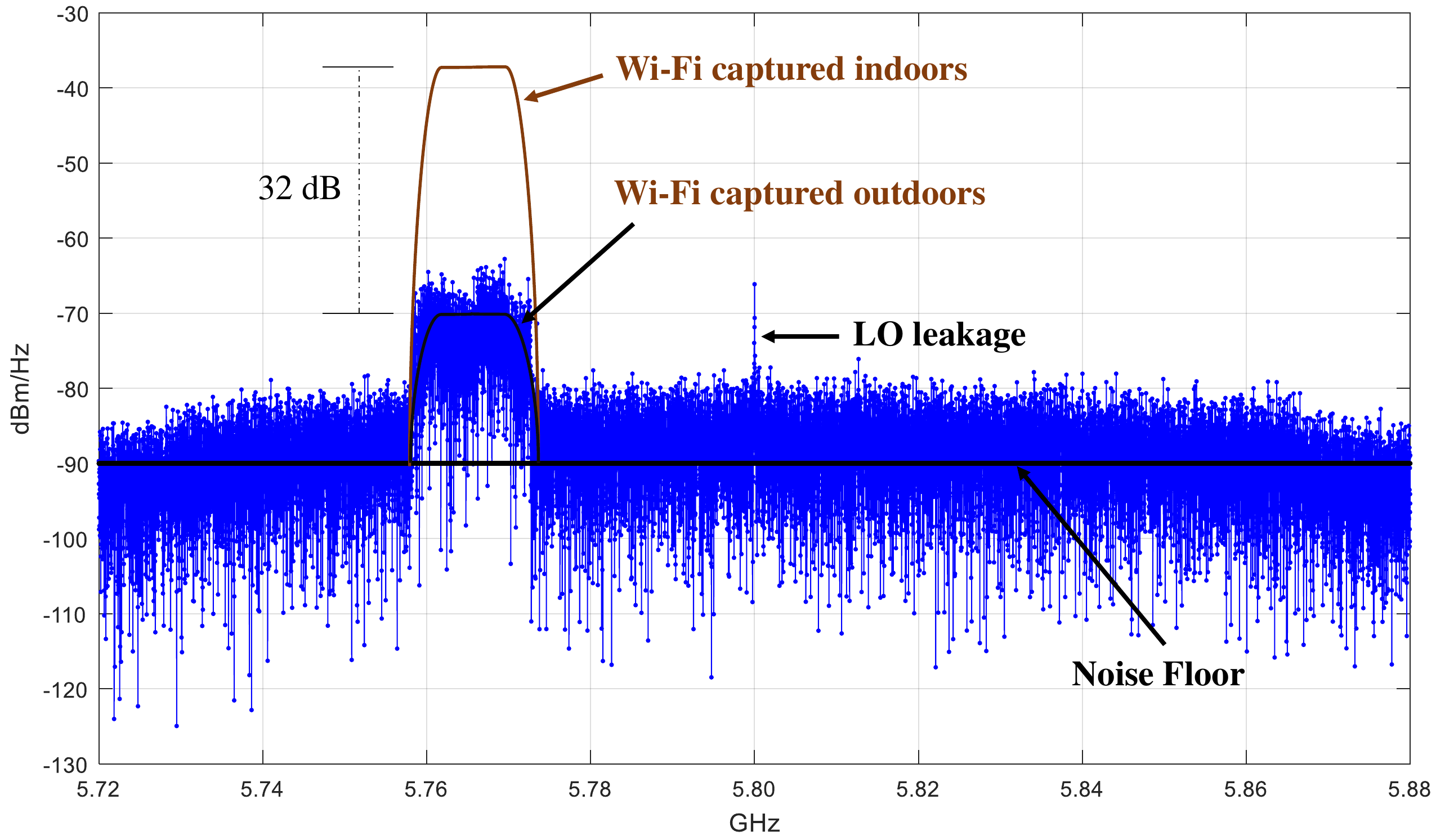}\\
  \caption{The PSDs of the 5.8 GHz Wi-Fi signal captured indoors and outdoors, wherein the AP transmits $20$ MHz Wi-Fi signal on channel 153 (5.765 GHz). The SNR of the outdoor received signal is $20$ dB over the $20$ MHz channel 153, and is $11.3$ dB over the $150$ MHz ISM band.}
\label{Fig4}
\end{figure}

The Power Spectral Densities (PSDs) of the received signals captured indoors and outdoors are plotted in Fig. \ref{Fig4}. As can be seen, there is a $32$-dB gap between them. In particular, for the outdoor signal, the signal-to-noise ratio (SNR) is about $11.3$ dB over the $150$ MHz ISM band.

\begin{rem}
The PIS and SIS designed in this paper are spread spectrum signals over the $150$ MHz ISM band.
As will be shown later, the interference from the Wi-Fi captured in the experiment is negligible for the designed PIS/SIS signal as far as misdetection
rate and false alarm rate are concerned.
\end{rem}

\subsubsection{Interrupt signal design}
This subsection pres\-ents the design of PIS and SIS on the $5.8$ GHz ISM band. We only explain the generation and detection of PIS in the following, SIS is generated and detected similarly.

The PIS consists of $Q$ $N$-point Zadoff-Chu (ZC) sequences \cite{chuseq} embedded in a $Q$-point maximum length sequence (m-sequence \cite{mSequence}), for a total of $QN$ samples.
Denote the m-sequence by $\bm{c_p}$. Let $\bm{z}$ be a ZC sequence given by
\begin{eqnarray}
\label{Equ1}
z[n]=
\begin{cases}
\exp{\frac{-j\pi J n(n+1)}{N}} & \text{for $N$ odd,}\\
\exp{\frac{-j\pi J n^2}{N}} & \text{for $N$ even,}
\end{cases},
\end{eqnarray}
where $n=0,1,2,...,N-1$, and $J$ is a positive integer coprime to $N$. For our application, we set $J=1$ (the reason for choosing $M=1$ will be explained later).

Then, the $QN$-point PIS $\bm{I_p}$ is generated by
\begin{eqnarray}
\label{Equ2}
\bm{I_p} = \bm{c_p} \otimes \bm{z},
\end{eqnarray}
where $\otimes$ is the Kronecker product, and each element in $\bm{I_p}$ is given by
\begin{eqnarray}
{I_p}[i]={c_p}[\left \lfloor i/N \right \rfloor]{z}[i \bmod N] \nonumber
\end{eqnarray}
for $i=0,1,2,...,QN-1$.
In \eqref{Equ2}, the ZC sequence acts like a spread spectrum sequence with rate $150$ MHz, thereby spreading the power of the m-sequence over the $150$ MHz band.

The receiver computes two cross-correlations \cite{APNC} to detect the PIS. Given the received sequence $\bm{r}$ (i.e., the $150$ MHz samples after ADC), the receiver first cross-correlates $\bm{r}$ and $\bm{z}$ as follows:
\begin{eqnarray}\label{Equ3}
y[i] = \sum_{n=0}^{N-1}z^*[n]r[i+n].
\end{eqnarray}
Note that the target interrupt signal is embedded in $\bm{r}$. Thus, in the presence of an interrupt signal, the operation in \eqref{Equ3} produces $Q$ peaks if we look at the absolute values of the resulting sequence $\bm{y}$, thanks to the correlation property of ZC sequences. Then, we make use of the m-sequence $\bm{c_p}$ modulated on the ZC sequence, and accumulate the power of all $Q$ peaks, yielding
\begin{eqnarray}\label{Equ4}
\overline{y}[i] = \left|\sum_{q=0}^{Q-1}c_p[q]y[i+Nq]\right|.
\end{eqnarray}
Finally, a sharp peak emerges at $\overline{y}[0]$. The capture of this peak results in successful detection of PIS.

For the SIS, the same ZC sequence is used, but in place of $\bm{c_p}$, another $Q$-point m-sequence $\bm{c_s}$ is used.

\begin{figure}[t]
  \centering
  \includegraphics[width=0.75\columnwidth]{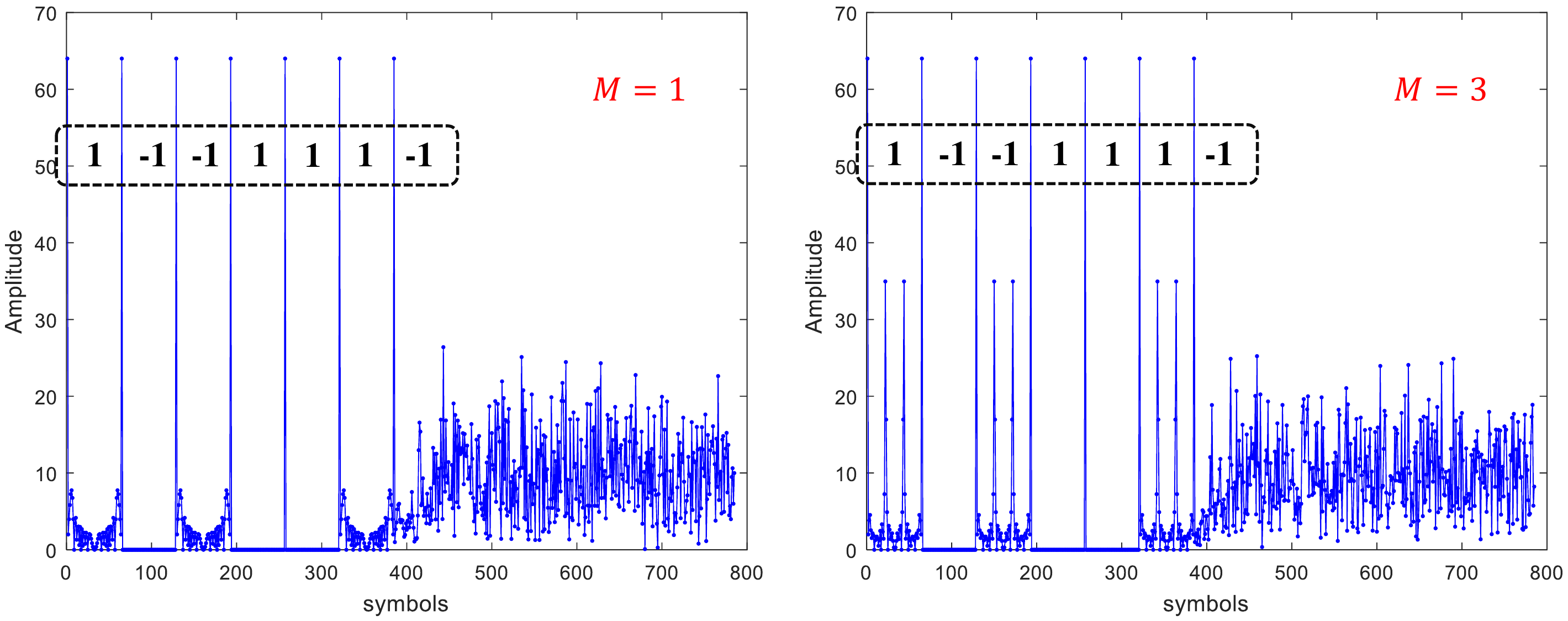}\\
  \caption{The amplitude of the cross-correlation results, i.e., $|\bm{y}|$. We set $N=64$, $Q=7$, $M=1$ or $3$. The orthogonality no longer holds when the adjacent two ZC sequences in PIS are modulated by opposite values.}
\label{Fig5}
\end{figure}

Before analyzing the detection performance, let us explain why $J$ is set to $1$ when generating the ZC sequence. ZC sequences have a nice correlation property: the periodic autocorrelation function of a ZC sequence is zero everywhere except at a single maximum per period \cite{CDMA}. However, when we modulate m-sequence onto ZC sequence, this nice correlation property no longer holds.
\begin{prop}
If these two adjacent ZC sequences are modulated by same values $1$ or $-1$, then $|y[\ell]|=0$ for $\ell=1,2,3,...,N-1$, because a ZC sequence is orthogonal to its cyclic shift. However, if two adjacent ZC sequences are modulated by opposite values $1$ and $-1$,
\begin{eqnarray}\label{Equ5}
\left |y[\ell]\right |=2\times\left |\frac{\sin(\pi M\ell^2/N)}{\sin(\pi M\ell/N)}\right |,
\end{eqnarray}
where $\ell=1,2,3,...,N-1$.
\end{prop}
\begin{NewProof}
See Appendix \ref{sec:appA}.
\end{NewProof}
As can be seen, when the two adjacent ZC sequences in PIS are modulated by opposite values, the resulting cross-correlated signal is in general nonzero at $l=1,2,3,...,N-1$.
The cross-correlated signals for $M=1$ and $M=3$ are shown in Fig. \ref{Fig5}.
Among all possible $M$, we found that setting $M\!=\!1$ minimizes the maximal interference $\max_l |y[l]|$ as well as the overall interference $\sum_l|y[l]|$. Thus, $M$ is set to $1$ to generate the ZC sequence.

\begin{rem}
If we use a long $NQ$-point ZC sequence instead of our design in this paper (i.e., $Q$ cascaded $N$-point ZC sequences), the detection performance will be the same. However, long ZC sequences greatly increases the computational complexity. Specifically, 1) for our design, the two-step cross-correlation takes only $N$ complex multiplications (other operations, e.g., complex additions, are negligible compared with the complex multiplication); 2) for a long $NQ$-point ZC sequence, however, one long $NQ$-point cross-correlation is needed, and it takes $NQ$ complex multiplications.
\end{rem}
\subsection{Neyman-Pearson Detector}\label{sec:NPdetector}
There are three components in the received sequence $\bm{r}$ (i.e., the $150$ MHz samples after ADC): the target interrupt signal, the $5.8$ GHz Wi-Fi signal as interference, and noise. Let us focus on the detection of PIS, giving
\begin{eqnarray}\label{EquDr}
\bm{r} = \sqrt{\rho_I} \bm{I_p} + \sqrt{\rho_x} \bm{I_x} + \bm{w},
\end{eqnarray}
where $\rho_I$ is the average received power of the PIS; $\bm{x}$ is the discrete samples of the Wi-Fi signal with an average received power of $\rho_x$; and $\bm{w}$ is discrete samples of circularly-symmetric Gaussian complex noise \cite{TseBook}, $w[i]\sim\mathcal{CN}(0,\sigma_w^2)$. The received signal to interference plus noise ratio (SINR) is
\begin{eqnarray}\label{eq:sinr}
\textup{SINR}=\frac{\rho_I}{\rho_x+\sigma_w^2}.
\end{eqnarray}

The receiver then performs the two cross-correlations in \eqref{Equ3} and \eqref{Equ4} to compute $\bm{\overline{y}}$. In particular, the decision statistic $\overline{y}[0]\triangleq u$ is given by
\begin{eqnarray}\label{EquDu}
u
\hspace{-0.5cm}&&= \left|\sum_{q=0}^{Q-1}c_p[q] \sum_{n=0}^{N-1}z^*[n]r[Nq+n]\right| \nonumber\\
\hspace{-0.5cm}&&= \left|\sqrt{\rho_I}NQ + \sum_{q=0}^{Q-1}c_p[q] \sum_{n=0}^{N-1}z^*[n]\left(\sqrt{\rho_x} {x}[Nq+n] + {w}[Nq+n]\right)\right|  \nonumber\\
\hspace{-0.5cm}&&\triangleq \left|\sqrt{\rho_I}NQ + \sqrt{\rho_x} x_c + w_c \right|.
\end{eqnarray}
where $x_c$ and $w_c$ are the cross-correlation results of Wi-Fi signal and noise, respectively. In particular, both $x_c$ and $w_c$ are complex Gaussian random variables, as specified in Lemma \ref{thm:2} and \ref{thm:3}.

\begin{lem}[Variance of $w_c$]\label{thm:2}
The cross-correlation results of AWGN noise $w_c$ is a complex Gaussian random variable, $w_c\sim\mathcal{CN}(0,NQ\sigma^2_w)$.
\end{lem}

\begin{NewProof}
See Appendix \ref{sec:appB}.
\end{NewProof}

\begin{lem}[Gaussian approximation of $x_c$]\label{thm:3}
The cross-correlation results of Wi-Fi signal $x_c$ can be approximated by a complex Gaussian random variable, i.e., $x_c\sim\mathcal{CN}(0,\sigma^2_x)$, where
\begin{eqnarray}\label{EquDvarx}
\sigma^2_x\approx \frac{Q}{M}\sum_{n=0}^{N-1}z^*[n]\sum_{n'=0}^{N-1}z^*[n']\sum_{k=0}^{M-1}z^*[n]\exp\left(j\frac{2\pi k(n-n')}{FM} \right).
\end{eqnarray}
\end{lem}

\begin{NewProof}
See Appendix \ref{sec:appB}.
\end{NewProof}

We use a statistical hypothesis test to analyze the MDR and FAR of PIS detection (i.e., the detection of $u$). Following \eqref{EquDu}, we define two hypotheses:
\begin{eqnarray*}
H_0: u \hspace{-0.5cm}&&= \left|\sqrt{\rho_x} x_c + w_c \right|, \\
H_1: u \hspace{-0.5cm}&&= \left|\sqrt{\rho_I}NQ + \sqrt{\rho_x} x_c + w_c \right|,
\end{eqnarray*}
Hypothesis $H_0$ is that there is no PIS and hypothesis $H_1$ is that there is a PIS. The MDR $P_M=P(H_0\mid H_1)$ and FAR $P_F=P(H_1\mid H_0)$.

The {\it a priori} probabilities of the two hypotheses, i.e., $P(H_0)$ and $P(H_1)$, are unknown, but we do know $P(H_1)\ll P(H_0)$ because of the sporadicity of warning messages.
In this context, we design the detector to be a Neyman-Pearson detector \cite{NPTest}. That is, for a tolerable FAR $P_F=\alpha$, the detector aims to minimize the MDR $P_M$.

As per the Neyman-Pearson Lemma \cite{NPTest}, such detector makes the decision by
\begin{eqnarray}\label{eq:rule1}
\lambda(u) = \frac{P(u|H_1)}{P(u|H_0)} \mathop{\lessgtr}_{H_1}^{H_0} \gamma,
\end{eqnarray}
where the threshold $\gamma$ is chosen such that the FAR
\begin{eqnarray*}
P_F = \int_{\gamma}^{\infty}p(\lambda|H_0)d\lambda=\alpha.
\end{eqnarray*}
In a nutshell, the Neyman-Pearson receiver makes the decision based on the likelihood ratio $\lambda(u)$. Given an observation $u$, if the likelihood ratio $\lambda(u)$ is smaller than the threshold $\gamma$, the receiver decides that $H_0$ is true; if $\lambda(u)$ is larger than the threshold $\gamma$, the receiver decides that $H_1$ is true.

\begin{thm}[Performance of the Neyman-Pearson detector]\label{thm:4}
Given an observation $u$, and a tolerable FAR $P_F=\alpha$, the optimal decision criteria that minimizes the MDR is given by
\begin{eqnarray}\label{eq:rule11}
u \mathop{\lessgtr}_{H_1}^{H_0} u^* = \sqrt{-2\sigma^2_u\ln \alpha}
\end{eqnarray}
where $\sigma^2_u = \rho_x\sigma^2_x+NQ\sigma^2_w$. The corresponding MDR is given by
\begin{eqnarray}\label{eq:PM1}
P_M = 1 - Q_1\left(\frac{\sqrt{\rho_I} NQ}{\sigma_u},\frac{u^*}{\sigma_u} \right).
\end{eqnarray}
where $Q_1$ is a Marcum Q-function \cite{TseBook}.
\end{thm}

\begin{NewProof}
See Appendix \ref{sec:appC}.
\end{NewProof}

Theorem \ref{thm:4} indicates that, instead of comparing the likelihood ratio $\lambda(u)$ with a threshold $\gamma$, it is equivalent to compare the received $u$ with a threshold $u^*$.
The optimal threshold $u^*$ is determined by the Wi-Fi signal power, the noise power, and the tolerable FAR.

On the other hand, the SIS has the same structure as the PIS. The only difference is that, in place of $c_p$, another $Q$-point m-sequence $c_s$ is used to accumulate the power of $Q$ ZC sequences. Thus, the detection of a single SIS has the same MDR and FAR as the PIS. In our interrupt protocol, a single PIS-broadcasting triggers multiple SIS-broadcasting. A two-hop neighbor of the emergency node detects the emergency successfully as long as one SIS is detected. Thus, given the same FAR $P'_F=\alpha$ and the corresponding $u^*$ computed from \eqref{eq:rule11}, the MDR of a two-hop neighbor of the emergency node is $P'_M=1-(P_M)^D$, where $D$ is the number of SISs that can be heard by this two-hop neighbor.

\section{Channel Access}\label{sec:V}
After interruption, the service channels are set aside for ultra-time-critical crowd messaging. The next problem is the channel access of multiple emergency nodes. Overall, we can summarize the problem as follows:
\begin{itemize}
\item There are $K$ (out of $K_\textup{all}$) active nodes. Typically, $K\in[0,30]$ and $K_\textup{all}\in[0,1000]$.
\item All the active nodes intend to transmit a message ($24$ Bytes) within $T=10-T_I$ ms, where $T_I$ is the time consumed by interrupts.
\item The available bandwidth is $40$ MHz. In practice, we may override only $20$ MHz so that the primary traffic of the service channels would not be clipped suddenly, and can still transmit on the other $20$ MHz channels.
\end{itemize}

Schedule-based channel access protocols, e.g., TDMA, FDMA, CDMA, OFDMA, requires pre-allocating orthogonal resources for the overall $K_\textup{all}$ nodes.
Let us take CDMA for instance. When operated with CDMA, all the nodes within two hops are pre-assigned different spread spectrum codes, e.g., pseudorandom noise (PN) codes, so that the spread spectrum signals from distinct nodes will not interfere with each other.
For one thing, a background coordinator must run in all time to guarantee all the nodes within two hops use different PN codes; for another, since there is no prior information on the potential transmitters, a receiving node must despread the received signal using all the potential PN codes (up to a few thousands).
This poses great challenges to the processing capacity of the receiver. In this context, random channel access protocols are preferable in our framework.

\subsection{Multi-replica ALOHA}\label{sec:V1}
A simple random-access protocol is ALOHA \cite{ALOHA}.
However, ALOHA requires ACK to inform the transmitter whether the previous transmission is successful or not. As stated in the introduction, ACK is not viable for the broadcast scenario, because each broadcast requires feedback from all one-hop neighbors, incurring excessive overhead when the network is dense, hence compromising the ability to meet the critical time constraint.

\begin{figure}[t]
  \centering
  \includegraphics[width=0.6\columnwidth]{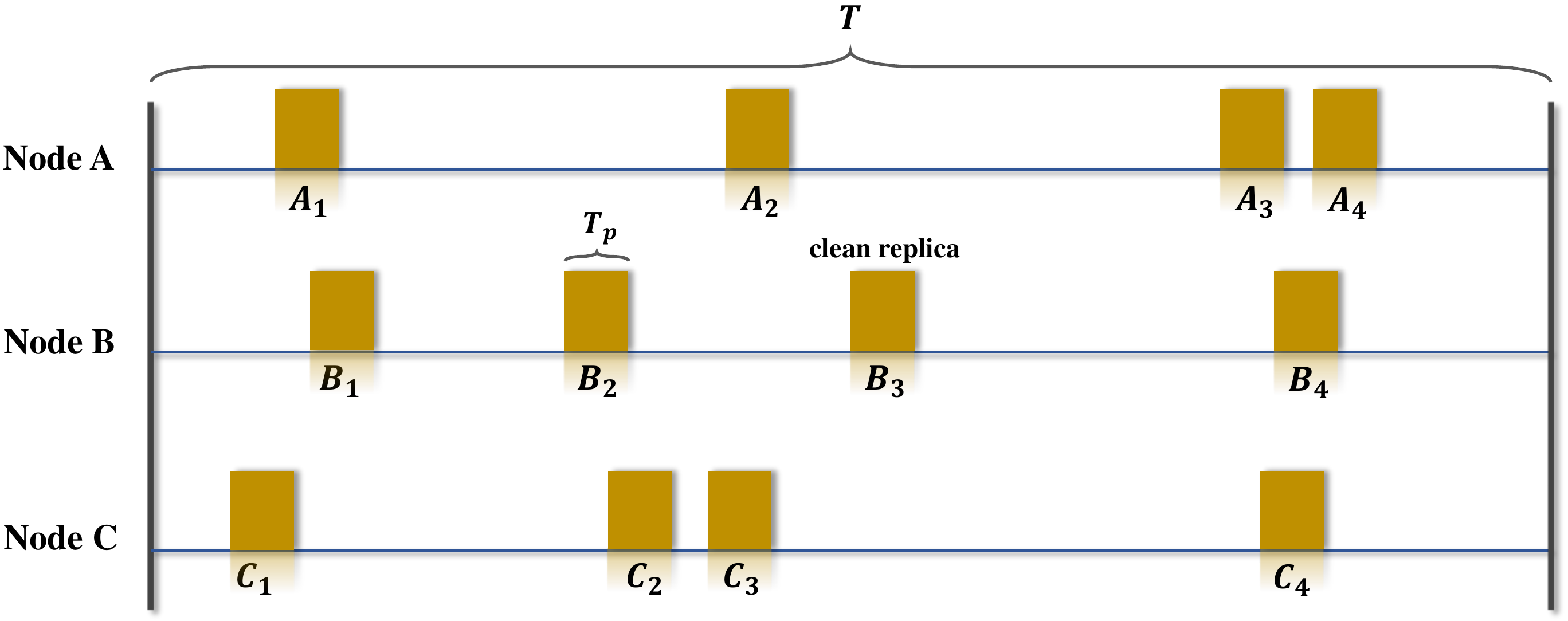}\\
  \caption{Multi-replica ALOHA. Each of the $K$ emergency nodes transmit $d$ replicas within $T$ ms. In this figure, $K=3$ and $d=4$. }
\label{Fig8}
\end{figure}

One alternative is multi-replica ALOHA.
The basic idea is that, since transmitters cannot determine whether their transmissions are successful or not given the lack of ACK, they can replicate their warning packet $d$ times and randomly broadcasts these $d$ replicas within $T$ ms. If one or more replicas from a node are broadcasted without any collision, then the delivery of the warning message is considered successful. An example is given in Fig. \ref{Fig8}, in which $d=4$, and three transmitters $A$, $B$ and $C$ broadcast four replicas, respectively. In this example, only replica $B_3$ is clean (i.e., no packet collides with $B_3$). Thus, only node $B$ successfully broadcasts its warning message while nodes $A$ and $C$ fail.

\begin{thm}[Message loss rate of multi-replica ALOHA]\label{thm:5}
When operated with multi-replica ALOHA, the message loss rate $R_{\textup{loss}}$, i.e., the probability that all d replicas of a node fail to be transmitted successfully, is given by
\begin{eqnarray}\label{eq:R_loss}
R_{\textup{loss}}\approx(1-P_0^{K-1})^d.
\end{eqnarray}
where $P_0$ is defined as the probability that a particular node $A$'s replica, say $A_1$, does not collide with any other node $B$'s $d$ replicas, giving
\begin{eqnarray}\label{eq:P0}
P_0=\frac{[T-(d+1)T_p]^{d+1}}{(T-dT_p)^d(T-T_p)},
\end{eqnarray}
and $T_p$ is the duration of one replica.
\end{thm}

\begin{NewProof}
See Appendix \ref{sec:appD}.
\end{NewProof}

The approximation in \eqref{eq:R_loss} comes from the assumption that the collision events of the $d$ replicas of a node are independent.
The approximation is valid to the extent that the time occupied by the $d$ replicas, $dT_p$, is much smaller than $T$, the total available time. This is the case for the scenarios of interest to us: we are interested in the low $R_{\textup{loss}}$ regime, and to have low $R_{\textup{loss}}$, a precondition would be that the available time $T$ is not crowded with replicas from the $K$ nodes.
Moreover, this is a conservative approximation in that the actual $R_{\textup{loss}}$ will be smaller than that given in \eqref{eq:R_loss}. The reason is that, if a replica experiences a collision, the probability of another replica experiencing a collision will be smaller than the probability of collision of the first replica. The collision of the first replica means that the replicas of the other nodes are more likely to overlap with the first replica, and therefore less likely to overlap with the second replica, since the first and second replicas of node $A$ do not overlap.

Given the number of active nodes $K$, we are interested in deriving the optimal degree $d^*$ to minimize the $R_{\textup{loss}}$.
However, it is tricky to compute the derivative of $R_{\textup{loss}}$ with respect to $d$ directly from \eqref{eq:R_loss}.
In this context, we resort to a Poisson approximation \cite{IFDMA} in Appendix \ref{sec:appE} to derive the optimal duplication factor $d^*$.

\begin{thm}[The optimal degree and the maximal number of sustainable nodes]\label{thm:6}
Given the number of active nodes $K$, the optimal $d^*$ in multi-replica ALOHA that minimizes the message loss rate $R_{\textup{loss}}$ is given by
\begin{eqnarray}\label{eq:dstar}
d^* \approx \frac{\ln 2}{2(K-1)}\frac{T}{T_p}.
\end{eqnarray}
Given a target message loss rate $R_{\textup{loss}}$, the maximal number of sustainable nodes in multi-replica ALOHA can be approximated by
\begin{eqnarray}\label{eq:Kstar}
K^*\approx -\frac{T}{T_p}\frac{\ln^2 2}{2\ln R_{\textup{loss}}}+1.
\end{eqnarray}
\end{thm}

\begin{NewProof}
See Appendix \ref{sec:appE}.
\end{NewProof}

Appendix~\ref{sec:appE} also draws an analogy between classical ALOHA and our problem using Multi-replica ALOHA.
In classical ALOHA, the transmission attempt rate $G$ involves the offered load (i.e., new arrivals) and the retransmissions.
The optimal $G$ is given by $G^*=\frac{1}{2}$ for unslotted ALOHA, in which case the optimal throughput is $G^*e^{-2G^*}=\frac{1}{2e}$.
In our problem, however, the objective is to lower the loss probability rather than maximizing the throughput.
In particular, the offered load in our problem is fixed to $\frac{KT_p}{T}$, and we allow $d$ attempts per node to jack up the transmission attempt rate to lower the loss probability.
The effective $G$ is therefore $\frac{KT_pd}{T}$ (i.e., number of attempts per packet duration).
Assuming large $K$ so that $K-1\approx K$, equation \eqref{eq:dstar} and $G=\frac{dKT_p}{T}$ imply that we have to achieve an effective $G=\frac{\ln 2}{2}$ to obtain the minimum message loss rate.
Note that the expression $G=\frac{\ln 2}{2}$ is independent of $d$ and $K$ under the adoption of the optimal $d$ for the given $K$.
This expression in turn implies that we need to modify the optimal transmission attempt rate of classical ALOHA, $G=\frac{1}{2}$, by a factor of $\ln 2=0.693$ in order to arrive at $G=\frac{\ln 2}{2}$ as  the optimal transmission attempt rate for our problem.

Overall, multi-replica ALOHA is a simple channel-access technique. Compared with other advanced techniques (e.g., the coded ALOHA introduced below), simplicity is its most attractive property. In particular, the signal processing of multi-replica ALOHA will not consume much additional time. Thus, for a target $R_\textup{loss}$, if the number of transmitters $K$ is no more than $K^*$ given in \eqref{eq:Kstar}, we would recommend Multi-replica ALOHA for reliable channel access with message loss rate less than $R_\textup{loss}$.
On the other hand, for the target $R_\textup{loss}$, if $K$ exceeds $K^*$ in (10), we need to resort to more advanced techniques using more complex signal processing. In subsection~\ref{sec:codedAloha} below, we explore the use of coded ALOHA to increase the sustainable $K$.

\subsection{Coded ALOHA}\label{sec:codedAloha}
At the transmitter, as with the multi-replica approach, each emergency node repeats its broadcast for $d$ times to increase the success rate. At the receiver, successive interference cancellation (SIC) \cite{SIC} can be used to boost performance.
In this paper, by Coded ALOHA, we mean that the SIC technique is used at the receiver to extract messages. This includes the same set-up that we studied in subsection~\ref{sec:V1} where a transmitter just repeats its message $d$ time (i.e., repetitive code is used), with the difference that SIC is used at the receiver to reduce message loss rate.

Consider the example in Fig. \ref{Fig8} again. Only $B_3$ can be decoded with the previous multi-replica reception mechanism.
With coded ALOHA, the receiver stores all the signals received during the $T$ ms, and make use of SIC to recursively cancel the interference caused by the decoded nodes.
First, the clean replica $B_3$ can be used to cancel other replicas of node $B$, i.e., $B_1$, $B_2$, and $B_4$.
As a result, the interference from node $B$ to other nodes is removed.
Moreover, this interference cancellation process creates a new clean replica $C_2$, and all node $C$'s replicas can be removed accordingly.
Finally, only replicas from node $A$ is left, and they are all clean and decodable.

This scheme, multi-replica ALOHA with SIC (or for simplicity, coded ALOHA), is similar to coded slotted ALOHA \cite{CSA,NCSA}, except for the absence of the concept of slotted time in the former.
Practically, a time-slotted system causes two problems in our application:
1) The slot must be short, e.g., as short as $24~\mu s$. However, small slot duration means larger overhead on slot alignment/synchronization among nodes.
The required guard time between slots will eat up a large portion of the slot time.
2) Alignment and synchronization of slots in the slotted system must be maintained all the time since we cannot predict the arrival of emergencies.
That is, nodes will need to participate in the slot synchronization process whether they currently have urgent messages to transmit, in preparation for possible arrivals of urgent messages -- performing synchronization only after the arrivals of messages will likely to cause unacceptable latency.

\textbf{The optimal degree distribution} -- The performance of coded (unslotted) ALOHA is analytically intractable due to the lack of mathematical tools to characterize the embedded SIC process.
On the other hand, in coded slotted ALOHA, the SIC decoding process can be analytically described by iterative message passing (i.e., the evolution of the erasure probabilities) on a bipartite graph \cite{IRSA,CSA}.
The bipartite graph consists of Burst Nodes (BNs), Sum Nodes (SNs) and edges. For example, a BN is a warning message transmitter, a SN is a slot, and an edge connects a BN and a SN if and only if a replica of the BN is transmitted in the SN/slot. The number of edges connected to a BN is referred to as the BN degree. Graphs for which the BN degree is constant are referred to as {\it regular} graphs; otherwise, the graphs are referred to as {\it irregular} graphs.

An important insight from the graph analyses of coded slotted ALOHA is that, the optimal throughput performance is often achieved by irregular graphs, whereas regular graphs usually lead to a performance loss \cite{LDPC,IRSA}.

For our problem, instead of using a fixed degree (transmitting a fixed number of replicas), we let each emergency node sample a degree from a degree distribution $\{\lambda_1,\allowbreak\lambda_2,\allowbreak\cdots,\allowbreak\lambda_d,\allowbreak\cdots,\allowbreak\lambda_D\}$, where $\lambda_d$ is the probability that the node chooses degree $d$.
The polynomial representation of the degree distributions is given by
\begin{eqnarray}
\Lambda_D(x)=\sum_{d=1}^{D}\lambda_d x^d.
\end{eqnarray}

The problem is then to discover the optimal degree distribution $\Lambda^*_{D^*}(x)$ to minimize the message loss rate $R_{\textup{loss}}$.
As will be shown in section \ref{sec:VIC}, the regular distribution $\Lambda_4(x)=x^4$ has already met the reliability requirements of warning messages. However, for the problem itself, the optimal degree distribution $\Lambda^*_{D^*}(x)$ is yet unknown due to the lack of mathematical tools for coded unslotted ALOHA.

We note that our problem is different from the problem studied previously in the context of coded slotted ALOHA \cite{IRSA}. The most obvious difference is that ours is an unslotted system while \cite{IRSA} studied a slotted system. A more subtle difference is that the degree distribution obtained in \cite{IRSA} is one that optimizes the throughput in the asymptotic limit when the number of active nodes $K$ goes to infinity. For our problem set-up, $K$ is finite, and the offered load $KT_p/T$ (therefore the target throughput) is low. For a given $K$ and offered load $KT_p/T$, our problem is to find the optimal degree distribution that minimizes $R_\textup{loss}$. For example, with $T_p/T=24/9500$ and $K =30$, the offered load is only $0.076$. In essence, we are trying to achieve low latency (small $T$) and high reliability (low $R_\textup{loss}$) with a finite node population (finite $K$); whereas in \cite{IRSA}, the aim is to study the asymptotic throughput in the limit that $K$ (and therefore $T$) goes to infinity. Because of these fundamental differences, it is not clear that the degree distribution optimal for the problem set-up in \cite{IRSA} is also optimal in the context of high reliability with low latency such as in our problem set-up. The simulation results in section \ref{sec:VIC} shows that the answer in no (see Table~\ref{Table2}).

\section{Numerical and Simulation Results}\label{sec:VI}
\subsection{Wireless Interrupt}
In case of emergency, an interrupt node A broadcasts a PIS, and this PIS triggers multiple SISs by one-hop neighbors of A.
For node A's one-hop neighbors, detection of the PIS peak means a successful interrupt; for node A's two-hop neighbors, detection of at least one SIS peak means a successful interrupt.
Thus, SIS detection has better performance (smaller MDR) than the PIS detection because a two-hop neighbor has more chances to capture the peak.
In other words, PIS detection is the bottleneck of our wireless interrupt protocol.
This subsection evaluates the performance of PIS detection using the Neyman-Pearson detector devised in Section \ref{sec:NPdetector}.

Following our experimental results in Fig.~\ref{Fig4}, we set the noise floor to $-90$ dBm/Hz, i.e., the noise power $\sigma^2_w=\allowbreak 1.5\times\allowbreak 10^{-4}~W$. The received power of PIS is set to equal the noise power, i.e., $\rho_I=\allowbreak 1.5\times\allowbreak 10^{-4}~W$ (that is, we artificially let the noise drown out the PIS). Then, we set different power of Wi-Fi signal, i.e., $\rho_x$, to vary the SINR, as in \eqref{eq:sinr}.

The tolerable FAR $P_F$ is set to $10^{-7}$. FAR is the probability that we detect a false alarm within a sample sequence of PIS length (i.e., $QN$ samples). Thus, the number of false alarms per hour can be calculated by $P_F\times 3600\times 150$ MHz$/QN$. Let $N = 1024$ and $Q = 63$, for example, the number of false alarms per hour is only $0.83$ given $P_F=10^{-7}$.

\begin{figure}[t]
  \centering
  \includegraphics[width=0.5\columnwidth]{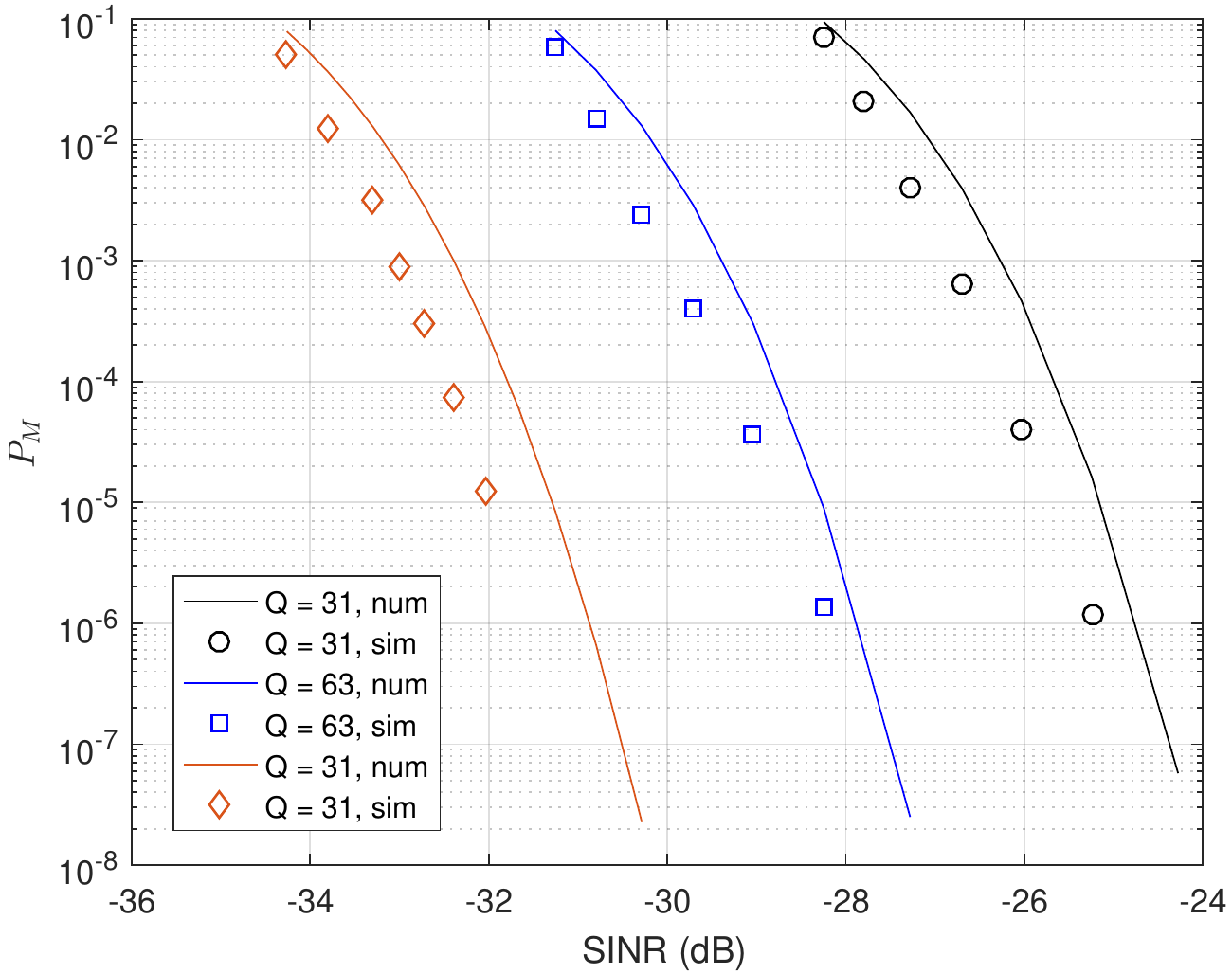}\\
  \caption{Numerical and simulation results of MDR versus SINR (in dB), wherein $N=1024$, $Q=31$, $63$ and $127$.}
\label{fig:MDR}
\end{figure}

The performance of MDR versus SINR (in dB) is shown in Fig. \ref{fig:MDR}, wherein we set $N = 1024$, $Q = 31$, $63$, and $127$ to generate the ZC sequence and m-sequence, respectively. The numerical results are plotted as per equation \eqref{eq:PM1}.

Two observations from Fig. \ref{fig:MDR} are as follows.
\begin{enumerate}
\item The analytical results are worse than the simulation results for about $0.5$ dB. This performance gap is caused by the approximation in \eqref{EquDvarx}, where we overcount the variance of the Wi-Fi signal to simplify the computations (see Appendix \ref{sec:appB} for more details).
\item The detection performance is getting better as $Q$ increases. This matches our intuition because longer sequence yields larger peaks. Let $P_M=10^{-6}$ be the target MDR, then setting $Q=31$ can meet the requirement when SINR $\geq -25.4$ dB; setting $Q=63$ can meet the requirement when SINR $\geq -28.2$ dB; setting $Q=127$ can meet the requirement when SINR $\geq -31.6$ dB. Moreover, when $Q=63$, PIS is composed of $NQ=64512$ symbols with a duration of  $0.43$ ms. If we reserve $0.07$ ms\footnote{For an FPGA chip with a 100 MHz clock rate, $0.07$ ms means $700$ clock periods. If we deploy two complex multipliers in parallel, $700$ clock periods are sufficient to compute the $N = 1024$ complex multiplications. Moreover, the propagation delay from the transmitter to the receiver is negligible, because the V2X communication range is usually less than $400$ meters, this corresponds to $1\sim 2~\mu$s propagation delay.} for signal processing, wireless interrupt consumes $T_I=0.5$ ms in total.
\end{enumerate}

\subsection{Multiple Access: Multi-replica ALOHA}
This subsection verifies the performance of the multi-replica ALOHA. In particular, we are interested in the message loss rate $R_\textup{loss}$, the optimal duplication factor $d^*$ given a fixed number of active nodes $K$, and the maximal number of sustainable nodes given a target message loss rate.

To these ends, we consider a specific OFDM-based PHY layer, the parameters of which are given in Table~\ref{Table1} \cite{DSRC}.
In particular, 1) the available bandwidth is $20$ MHz. That is, the warning messages will override half of the service channels so that the non-safety messages would not be totally deprived of services. 2) A typical $24$ Byte warning message occupies two OFDM symbols, leading to a $24~\mu s$ warning packet at the PHY layer (each OFDM symbol is $8~\mu s$ and the preamble is $8~\mu s$). 3) The time for interruption is $T_I=0.5$ ms, hence, the available time for channel access is $T=9.5$ ms.

\begin{table}[t]
\renewcommand{\arraystretch}{0.7}
\centering
\caption{}\label{Table1}
\setlength{\tabcolsep}{10mm}{
\begin{tabular}{ccc}
\toprule
Types                                                                       & Description                                            & Value          \\ \midrule
\multirow{8}{*}{\begin{tabular}[c]{@{}c@{}}OFDM\\ PHY\end{tabular}}            & available  bandwidth                                    & $20$ MHz       \\
                                                                            & subcarrier spacing                                     & $156.25$ KHz   \\
                                                                            & available data subcarriers                             & $96$           \\
                                                                            & modulation                                             & QPSK           \\
                                                                            & channel code rate                                      & $1/2$          \\
                                                                            & CP duration                                            & $1.6~\mu s$    \\
                                                                            & OFDM symbol duration                                   & $8~\mu s$      \\
                                                                            & preamble duration                                      & $8~\mu s$      \\\midrule
\multirow{8}{*}{\begin{tabular}[c]{@{}c@{}}{\it warning}\\ message\end{tabular}}& potential transmitters                                 & $K\in[0,30]$   \\
                                                                            & {\it warning} message size                             & $24$ Bytes     \\
                                                                            & {\it warning} packet duration                          & $24~\mu s$     \\
                                                                            & Overall TTL of {\it warning} messages                  & $10$ ms        \\
                                                                            & Time consumed by interruption                          & $T_I =0.5$ ms  \\
                                                                            & Time left for channel access                           & $T = 9.5$ ms \\

\bottomrule
\end{tabular}
}
\end{table}

\begin{figure}[t]
  \centering
  \includegraphics[width=0.52\columnwidth]{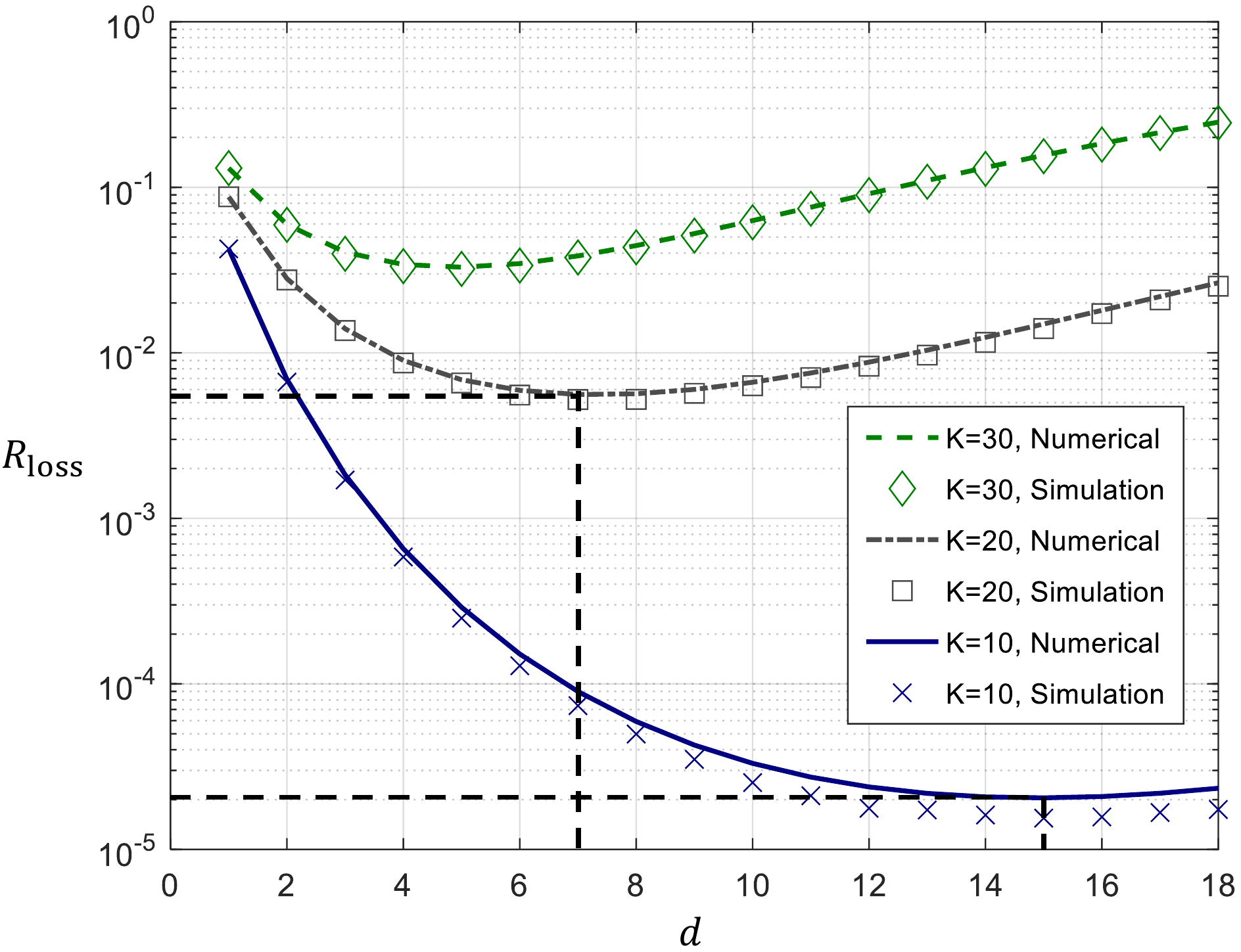}\\
  \caption{The message loss rate $R_{\textup{loss}}$ of multi-replica ALOHA, where $K=10$, $20$, and $30$. Each of the $K$ emergency nodes broadcasts $d$ replicas of their warning packets. The duration of a replica is $T_p=24~\mu s$, the overall time available for channel access is $T=9.5$ ms. The results presented here are based on \eqref{eq:R_loss} and simulations. The numerical and simulation results match well.}
\label{Fig9}
\end{figure}

\begin{figure}[t]
  \centering
  \includegraphics[width=0.52\columnwidth]{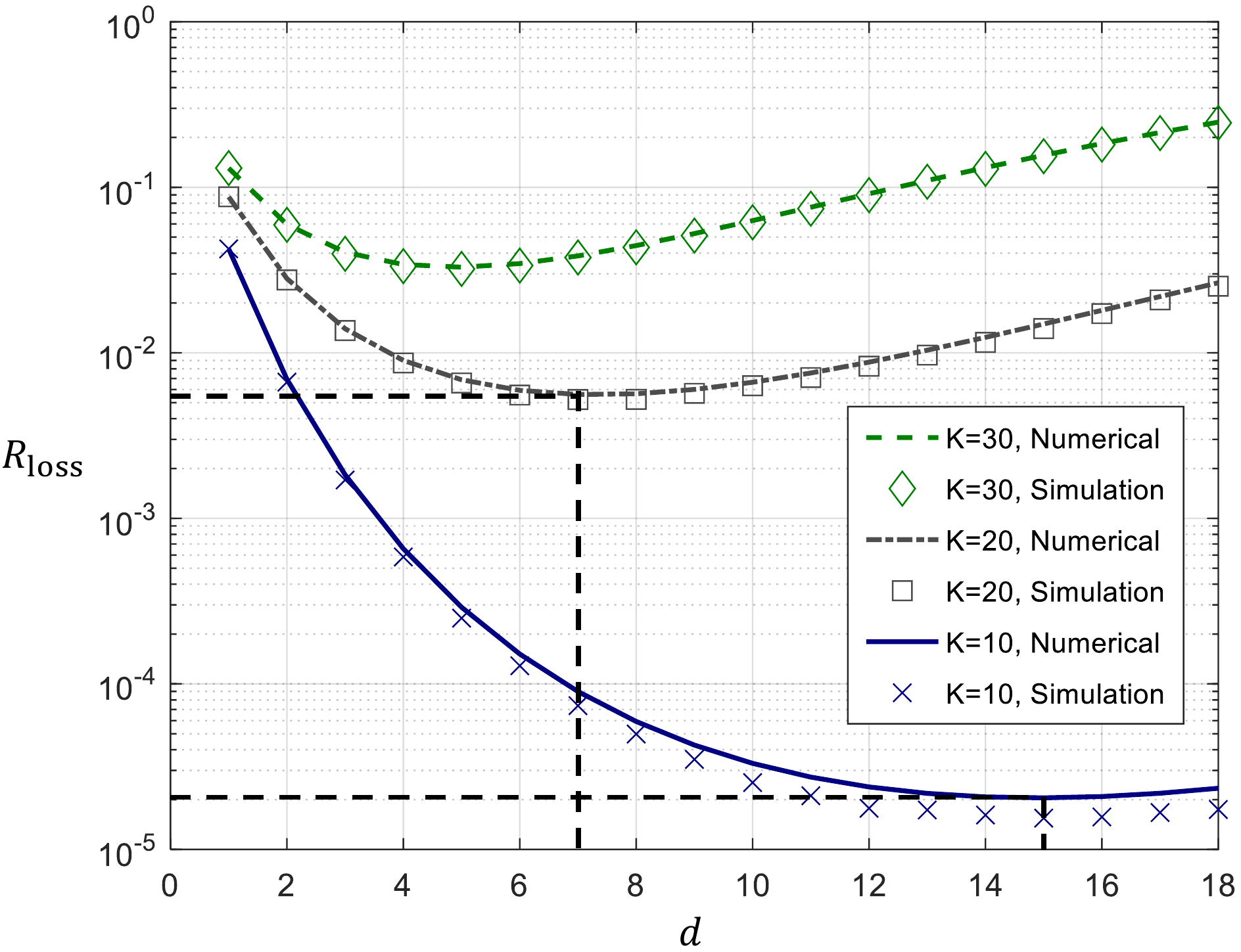}\\
  \caption{The message loss rate $R_{\textup{loss}}$ of multi-replica ALOHA, where $K=11$, $12$. Each of the $K$ emergency nodes broadcasts $d$ replicas of their warning packets. The duration of a replica is $T_p=24~\mu s$, the overall time available for channel access is $T=9.5$ ms. The results presented here are based on \eqref{eq:R_loss} and simulations. The maximum number of sustainable nodes in the system is $11$ if $R_\textup{loss}$ is to be no more than $10^{-4}$.}
\label{Fig9b}
\end{figure}

First, we focus on the message loss rate $R_\textup{loss}$. As shown in Fig.~\ref{Fig9} and \ref{Fig9b}, numerical and simulation results of $R_\textup{loss}$ are plotted in the same figure, wherein we set $K=\{10,20,30\}$ and $K=\{11,12\}$, respectively. The numerical results are computed as per equation \eqref{eq:R_loss}.
As shown, the numerical and simulation results match well, with the $R_\textup{loss}$ obtained by simulation slightly smaller than the $R_\textup{loss}$ obtained by \eqref{eq:R_loss}. This validates our earlier statement that the approximated \eqref{eq:R_loss} is good in the regime of our interest and the approximation tends to be on the conservative side.

Second, given a fixed number of active nodes $K$, we study the optimal duplication factor $d^*$ that minimizes the $R_{\textup{loss}}$. As can be seen, the approximate analytical expression in \eqref{eq:dstar} gives the right ballpark of the empirical optimal $d^*$ shown in Fig.~\ref{Fig9} and \ref{Fig9b}.
For example, when $K=10$, the predicted optimal degree is $15.24$, while the simulated $d^*=15$; when $K=20$, the predicted optimal degree is $7.22$, while the simulated $d^*=7$.

Finally, given $R_{\textup{loss}} = 10^{-4}$ as the target performance, the maximal number of sustainable nodes in multi-replica ALOHA is $K^*=11$, as shown in Fig.~\ref{Fig9b}. This matches with our prediction in \eqref{eq:Kstar} where the predicted $K^*=11.32$.

\subsection{Multiple Access -- Coded ALOHA}\label{sec:VIC}
This subsection evaluates the performance of coded ALOHA. We use the same PHY-layer parameters as in Table~\ref{Table1}, and assume perfect interference cancellation at the PHY layer.
\begin{figure}[t]
  \centering
  \includegraphics[width=0.5\columnwidth]{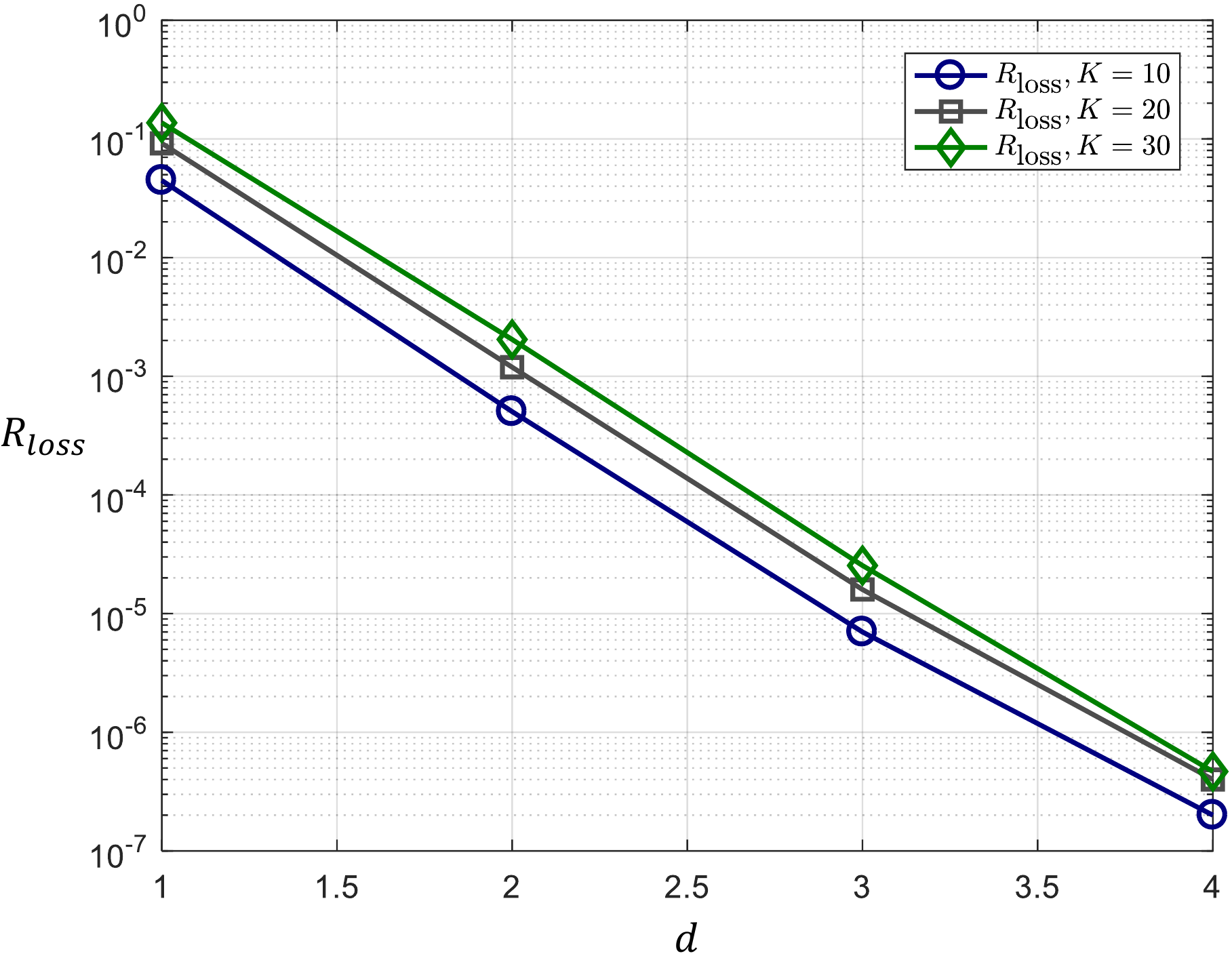}\\
  \caption{The message loss rate of coded ALOHA under different $d$, where the interference cancellation at the PHY layer is assumed to be perfect. The results presented here are based on simulations.}
\label{Fig10}
\end{figure}

The simulation results  are presented in Fig. \ref{Fig10}, where we simulated the message loss rate $R_{\textup{loss}}$ of coded ALOHA given different number of active nodes $K= 10$, $20$, and $30$. As can be seen, when $d\geq 3$, the $10^{-4}$ requirement is satisfied for all $K\leq 30$.

\begin{figure}[t]
  \centering
  \includegraphics[width=0.48\columnwidth]{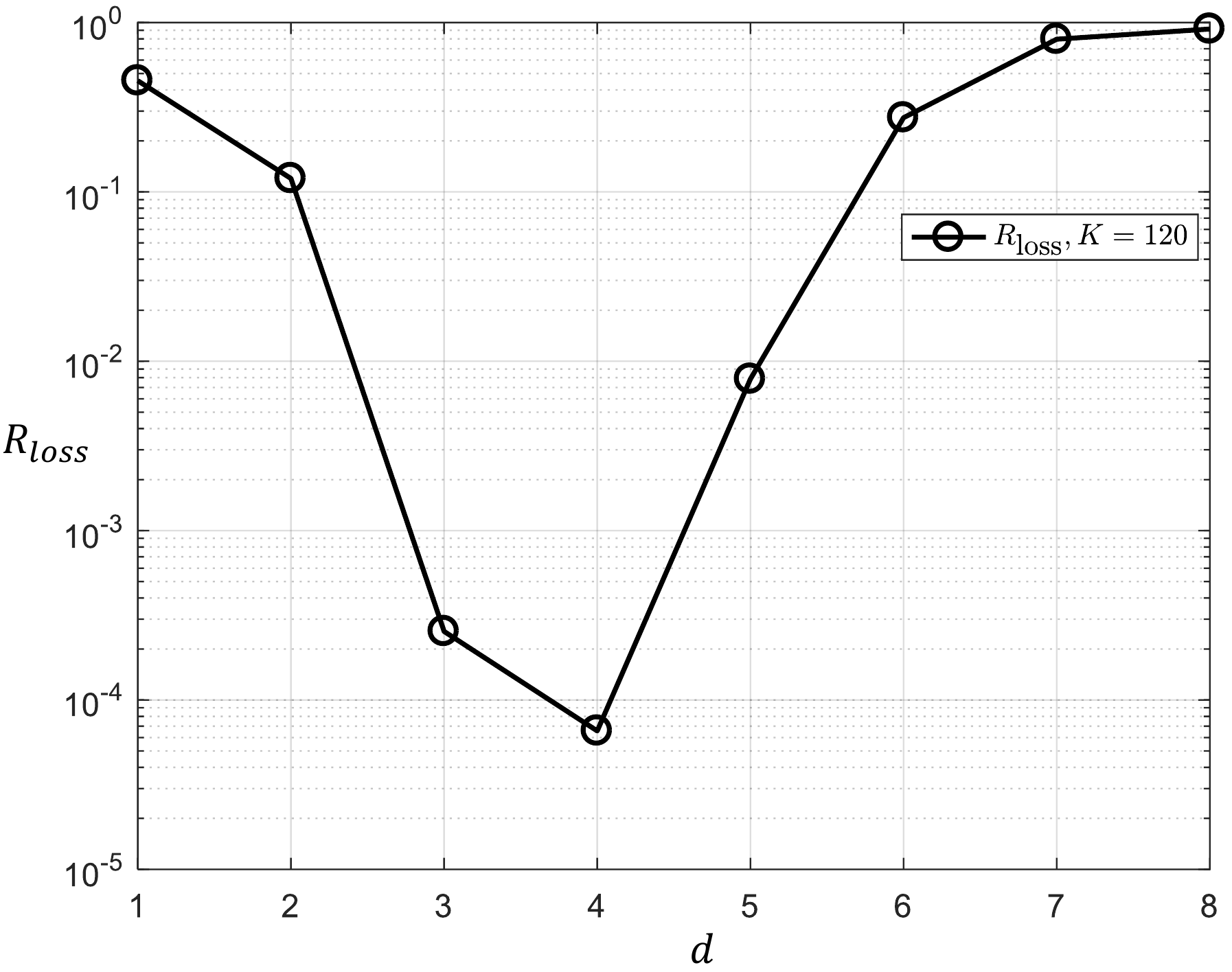}\\
  \caption{The message loss rate and global loss rate of coded ALOHA under different $d$, where $K=120$.}
\label{Fig10b}
\end{figure}

To measure the maximal number of sustainable nodes $K^*$ in coded ALOHA systems, we keep increasing the number of active nodes $K$ and simulate $R_\textup{loss}$ given different degree $d$. Fig.~\ref{Fig10b} shows the $R_\textup{loss}$ when $K=120$.
Approximately, the maximal number of sustainable nodes in coded ALOHA systems is $K^*=120$ given a target performance $R_{\textup{loss}} = 10^{-4}$.


\begin{table}[t]
\centering
\caption{Performance of different degree distributions, $K=30$.}\label{Table2}
\begin{tabular}{ccc}
\toprule
Distributions & $G^*$ in CSA     & $R_{\textup{loss}}$  \\ \midrule
$0.5102x^2+0.4898x^4$            & $0.868$     &  $4.5437\times 10^{-4}$       \\
$0.5x^2+0.28x^3+0.22x^8$         & $0.938$     &  $4.3776\times 10^{-4}$       \\
\begin{tabular}[c]{@{}l@{}}$0.4977x^2+0.2207x^3+0.0381x^4+$  $0.0756x^5+0.0398x^6+0.0009x^7+$\\ $0.0088x^8+0.0068x^9+0.0030x^{11}+$  $0.0429x^{14}+0.0081x^{15}+0.0576x^{16}$\end{tabular}
                                 & $0.965$     &  $4.3449\times 10^{-4}$       \\\midrule
$x^3$                            &             &  $2.5399\times 10^{-5}$       \\
$x^4$                            &             &  $4.6666\times 10^{-7}$      \\ \bottomrule
\end{tabular}
\end{table}

We performed additional simulations to verify if the optimal degree distributions designed for coded slotted ALOHA can be applied to our problem of coded unslotted ALOHA. The simulation results are presented in Table~\ref{Table2}, in which $K=30$. The first three rows of Table~\ref{Table2} are irregular distributions (with maximal degrees $4$, $8$, and $16$) designed for coded slotted ALOHA \cite{IRSA}. With greater maximal degree, higher asymptotic threshold of throughput, denoted by $G^*_0$, can be achieved (if the offered load is smaller than $G^*$, the messages can be recovered from the SIC process with a probability close to $1$ in the asymptomatic limit when $K\rightarrow\infty$). The last two rows in Table~\ref{Table2} are regular distributions used in Fig.~\ref{Fig10}.
As shown, the regular distributions outperforms irregular distributions by much.

In practice, the most time-consuming process in coded ALOHA is SIC. However, there is no need to wait until all replicas are received before the SIC begins. The receiver can perform the SIC while receiving the replicas, and this saves a lot of processing time.

\section{Conclusion}\label{sec:conclusion}
This paper studies the problem of life-critical warning messaging in V2X.
Our main contributions are as follows.
\begin{enumerate}
\item We put forth an interrupt-and-access MAC protocol for warning messaging that takes into account the three characteristics of warning messages: sporadicity, crowding, and ultra-time-criticality requirement. The idea is to interrupt the regular wireless services only when warning messages arrive so as to acquire usage of the spectrum ordinarily allocated to the regular services. In this way, precious wireless spectrum does not have to be pre-allocated to warning messaging, which occurs only once in a long while in a sporadic manner. The idea of interrupt-and-access coincides with the preemptive transmission in 5G NR-V2X proposed for urgent messages. Our solution is the first realization of the preemptive transmission scheme, to the best of our knowledge.
\item For wireless interrupt, we devised an interrupt mechanism for V2X and presented an out-of-band interrupt signal design where the interrupt signals are spread spectrum sequences on the ISM band. We analyzed the optimal Neyman-Pearson detector and derived the optimal detection threshold as well as the minimal misdetection rate (MDR). Numerical and simulation results validate the nice detection performance of our design, e.g., for a $0.43$ ms ($64512$ symbols, $150$ MHz) sequence, the MDR can be kept lower than $10^{-6}$ when SINR $\geq -28.2$ dB.
\item For wireless access, we investigated different uncoordinated channel access schemes to meet the stringent delay and reliability requirements of warning message. A simple multi-replica ALOHA scheme can support up to $11$ nodes in our set-up with message loss rate lower than $10^{-4}$. If the number of transmitters in the system exceeds $11$, a more advanced coded ALOHA scheme with successive interference cancellation can support up to $120$ nodes in our set-up while keeping the message loss rate lower than $10^{-4}$.
\end{enumerate}
\appendices
\section{Approximating the optimal degree $d^*$ and the maximal number of sustainable nodes $K^*$ in Multi-replica ALOHA}\label{sec:appE}
Consider the situation faced by one particular node, say, node $A$. On a line of length $T$, there are $(K-1)d$ points corresponding to the beginnings of the $(K-1)d$ replicas of the other $K-1$ nodes. We denote this set of points by $S$. To the extent that $K$ is large, the points in $S$ approximately form a Poisson process on the line. In other words, the inter-point distance is exponentially distributed with mean
\begin{eqnarray}
\mu=\frac{T}{(K-1)d}. \nonumber
\end{eqnarray}

\subsection{The optimal degree $d$ for a given $K$}
Consider a replica of node A, say $A_1$, that is randomly placed on the line of length $T$. Refer to the beginning of this packet as point $t(A_1)$. Ignoring the edge effects at the two ends of the line, the probability that the distance of point $t(A_1)$ to the next point of $S$ to the right is more than $T_p$ is $e^{-T_p/\mu}$. Similarly, the probability that the distance of point $t(A_1)$ to the next point of $S$ to the left is more than $T_p$ is $e^{-T_p/\mu}$. The probability of no collision is therefore $T_p$ is $e^{-2T_p/\mu}$. Thus, $R_{\textup{loss}}$ is approximately given by
\begin{eqnarray}\label{eq:R_loss_new}
R_{\textup{loss}} = (1-e^{-\rho d})^d,
\end{eqnarray}
where $\rho=\frac{2(K-1)T_p}{T}$. For the regime of our interest (i.e., $T\gg dT_p$, and $K\gg 1$), the $R_{\textup{loss}}$ in \eqref{eq:R_loss_new} is approximately equal to that in \eqref{eq:R_loss}. See Appendix \ref{sec:appF} for more details.

Differentiating $\ln R_{\textup{loss}}$ with respect to $d$ and setting the derivative to zero gives us the following equation:
\begin{eqnarray}
e^{-\rho d^*}\ln e^{-\rho d^*} = (1-e^{-\rho d^*})\ln (1-e^{-\rho d^*})\nonumber
\end{eqnarray}
This is satisfied by $e^{-\rho d^*}=1-e^{-\rho d^*}$, which gives
\begin{eqnarray}\label{eq:R_loss_equality}
e^{-\rho d^*}=\frac{1}{2}. \nonumber
\end{eqnarray}

The optimal $d^*$ is thus
\begin{eqnarray}\label{eq:optimald}
d^*=\frac{1}{\rho}\ln 2 = \frac{\ln 2}{2(K-1)}\frac{T}{T_p}.
\end{eqnarray}

If we look at Fig.~\ref{Fig9} wherein $\frac{T}{T_p}=\frac{9500}{24}$, we have
$d^*\approx \frac{137}{K-1}$,
which gives $d^*\approx 15.2$ for $K=10$;
$d^*\approx 7.2$ for $K=20$;
and
$d^*\approx 4.7$ for $K=30$.
This is consistent with the empirical optimal degree in Fig.~\ref{Fig9}.

More importantly, the above analysis reveals a fundamental relation between $K$ and $d$ in \eqref{eq:optimald}. That is, the optimal $d^*$ is inversely proportional to $K-1$.

\subsection{On the optimal transmission rate}
In classical ALOHA, $G$ is the the number of transmission attempts by all nodes per packet duration (including both the new arrivals and the retransmissions). The throughput of ALOHA is $Ge^{-2G}$ packets per packet duration. Thus, the optimal $G$ to maximize throughput is $G=\frac{1}{2}$, and the corresponding optimal throughput is  $Ge^{-2G}=\frac{1}{2e}$.
The study of classical ALOHA is to achieve this optimal throughput. If packets can be retransmitted indefinitely after back-offs until success, then there is no loss in the system as long as the offered load is less than $\frac{1}{2e}$ (subject to a proper backoff method). This may incur excessive delay, however.

In our problem, the offered load is fixed to $\frac{KT_p}{T}$, and we allow $d$ attempts per node. In other words, the effective $G$ in our problem is $\frac{KT_pd}{T}$.
Assuming large $K$ so that $K-1$ is approximately $K$, equation \eqref{eq:optimald} implies that we have to achieve $G=\frac{\ln 2}{2}$ to minimize the message loss rate. That is, we need to modify the optimal transmission attempt rate of classical Aloha, $G=\frac{1}{2}$, by a factor of $\ln 2=0.693$.

Note that, in our problem set-up, we do not adjust the offered load to try to meet the maximum throughput -- our offered load is already fixed (in fact smaller than the best sustainable offered load). We try to reduce the loss probability for a fixed offered load lower than the sustainable offered load of ALOHA, while bounding the delay to within $T$. The optimal $G$ will therefore be different.

\subsection{The maximal number of sustainable nodes $K^*$}
As far as one of the replicas is concerned, its success rate is given by $e^{-2G}=e^{-\ln 2}=\frac{1}{2}$ in the optimal setting -- i.e., half chance of success for each trial. Note that this success rate for a replica is independent of $K$ and $d$ because $K$ and $d$ have been optimized to give $G=\frac{\ln 2}{2}$. Thus, regardless of $K$, under the optimal setting, the failure rate after $d$ attempts of the $d$ replicas is $(\frac{1}{2})^d$. Of course, for a fixed $G$, the larger the $K$, the smaller the $d$.
For a given $K$, the minimum message loss rate is given by
\begin{eqnarray}\label{eq:optimalRloss}
R^*_{\textup{loss}}=\left(\frac{1}{2}\right)^d=\left(\frac{1}{2}\right)^{\frac{\ln 2}{2(K-1)}\frac{T}{T_p}}.
\end{eqnarray}
Eq.~\eqref{eq:optimalRloss} gives us an insight on how the minimum message loss rate $R^*_{\textup{loss}}$ depends on $K$ with the optimized $d$.
In the log scale, we have
\begin{eqnarray}
\ln R^*_{\textup{loss}}=-\frac{\ln^2 2}{2(K-1)}\frac{T}{T_p}. \nonumber
\end{eqnarray}

Given a target message loss rate $R_{\textup{loss}}$, the maximum $K^*$ is
\begin{eqnarray}
K^*=-\frac{T}{T_p}\frac{\ln^2 2}{2\ln R_{\textup{loss}}}+1.
\end{eqnarray}
For our settings where $R_{\textup{loss}}\leq 10^{-4}$, the maximum $K$ is $K^* = 11$. This is consistent with the numerical results in Fig.~\ref{Fig9}.

\bibliographystyle{IEEEtran}
\bibliography{References}

\newpage
\setcounter{page}{1}
\vspace{0.35in}

\begin{center}
{\large\bf Supplemental Materials for the Paper ``Sporadic Ultra-Time-Critical Crowd Messaging in V2X''}
\end{center}

\vspace{0in}

\section{Deriving the cross-correlation results}\label{sec:appA}
In this appendix, we derive the cross-correlation results when the two adjacent ZC sequences in PIS are modulated by distinct values.

First, from \eqref{Equ2}, the PIS is given by $\bm{I_p}=\bm{c_p}\otimes\bm{z}$, where $\bm{c_p}$ is a $Q$-point m-sequence and $\bm{z}$ is an $N$-point ZC sequence given by \eqref{Equ1}. In the following derivations, we consider even $N$ (odd $N$ yields the same results).

As with \eqref{Equ3}, at the receiver, we cross-correlate PIS $\bm{I_p}$ with the conjugate of ZC sequence $\bm{z}$, yielding
\begin{eqnarray}
y[i]=\sum_{j=0}^{N-1}z^*[j]I_p[i+j].\nonumber
\end{eqnarray}
Given sequence $\bm{y}$, we find peak in its modulus $|\bm{y}|$.
Without loss of generality, we now focus on the first two ZC sequences in PIS.

If these two ZC sequences are modulated by same values, then
\begin{eqnarray}\label{EquA1}
y[l] = \sum_{n=0}^{N-l-1}\!\!\!z^*[n]z[n\!+\!l]+\!\!\!\sum_{n=N-l}^{N-1}\!\!\!z^*[n]z[n\!-\!N\!+\!l] = 0,
\end{eqnarray}
where $l=1,2,3,...,N-1$. Eq.~\eqref{EquA1} follows since a ZC sequence is orthogonal to its cyclic shift.

If these two ZC sequences are modulated by distinct values, say $1$ and $-1$, respectively. We have
\begin{eqnarray}\label{EquA2}
y[l]
= \sum_{n=0}^{N-l-1}\!\! z^*[n]z[n\!+\!l]-\!\!\sum_{n=N-l}^{N-1}z^*[n]z[n\!-\!N\!+\!l]
= 2\sum_{n=0}^{N-l-1}z^*[n]z[n+l]
\end{eqnarray}

Substituting \eqref{Equ1} into \eqref{EquA2}, yields,
\begin{eqnarray}\label{EquA3}
\left |y[l]\right |=2\times\left |\frac{\alpha^{l^2}-\alpha^{-l^2}}{1-\alpha^{2l}}\right |,
\end{eqnarray}
where $\alpha=\exp{(-j\pi M/N)}$.

Notice that $\left |\alpha^{l^2}-\alpha^{-l^2}\right |$ and $\left |1-\alpha^{2l}\right |$ are two strings of a unit circle on the complex plane. According to the Law of cosines, we have
\begin{eqnarray}
\left |\alpha^{l^2}-\alpha^{-l^2}\right |
= \sqrt{1^2+1^2-2\cos(\frac{\pi M}{N}*2l^2)}
= \left |2\sin(\frac{\pi M}{N}l^2)\right |,\nonumber\\
\left |1\!-\!\alpha^{2l}\right |
= \sqrt{1^2+1^2-2\cos(\frac{\pi M}{N}*2l)}
= \left |2\sin(\frac{\pi M}{N}l)\right |,\nonumber
\end{eqnarray}

Thus, \eqref{EquA3} can be written as
\begin{eqnarray}
\left |y[l]\right |=2\times\left |\frac{\sin(\pi Ml^2/N)}{\sin(\pi Ml/N)}\right |.\nonumber
\end{eqnarray}

\section{Proof of Lemma \ref{thm:2} and \ref{thm:3}}\label{sec:appB}
In this appendix, we show both $w_c$ and $x_c$ are complex Gaussian random variables, and derive their variances.

\begin{NewProof}
From \eqref{EquDu}, we have
\begin{eqnarray*}
w_c = \sum_{q=0}^{Q-1}c_p[q] \sum_{n=0}^{N-1}z^*[n] {w}[Nq+n].
\end{eqnarray*}

Thus, $\mathbb{E}[w_c]=0$, and
\begin{eqnarray*}
\mathbb{E}[w_cw^*_c] \hspace{-0.5cm}&&= \mathbb{E}\Bigg[ \sum_{q=0}^{Q-1}c_p[q] \sum_{n=0}^{N-1}z^*[n] {w}[Nq+n]\sum_{q'=0}^{Q-1}c_p[q'] \sum_{n'=0}^{N-1}z^*[n'] {w}[Nq'+n']\Bigg]\\
\hspace{-0.5cm}&&= \sum_{q=0}^{Q-1}c^2_p[q] \sum_{n=0}^{N-1}z^*[n] \sigma^2_w \\
\hspace{-0.5cm}&&= NQ\sigma^2_w,
\end{eqnarray*}
As a result, $w_c\sim\mathcal{CN}(0,NQ\sigma^2_w)$.

Next, we show $w_c$ can be approximated as a complex Gaussian random variable.
The continuous $20$ MHz Wi-Fi signal can be written as
\begin{eqnarray*}
x(t) = \frac{1}{M}\sum_{k=0}^{M-1}s[k]\exp\left(j\frac{2\pi k t}{T_\textup{OFDM}}\right),
\end{eqnarray*}
where $M$ is the IFFT size, $T_\textup{OFDM}$ is the duration of one OFDM symbol, and $\{s[k]:k=0,1,2,\cdots,M-1\}$ are the frequency domain QAM-modulated symbols. In particular, $\{s[k]\}$ are independent and identically distributed (i.i.d.), and their mean and variance are given by
\begin{eqnarray}
&&\mathbb{E}[s[k]] = 0, \nonumber\\
\label{EquDvars}
&&\mathbb{E}[s[k]s^*[k']] = \sigma^2_s\delta(k-k'),
\end{eqnarray}

Sampling at $150$ MHz, the discrete Wi-Fi signal $\bm{x}$ at the receiver (before cross-correlation) is then
\begin{eqnarray}\label{eq:x150}
x[\ell] = \frac{1}{M}\sum_{k=0}^{M-1}s[k]\exp\left(j\frac{2\pi k \ell}{FM}\right),
\end{eqnarray}
where $F = 150 \textup{MHz}/ 20 \textup{MHz}=7.5$ is the oversampling rate. For large $M$, $x[\ell]$ can be approximated by a complex Gaussian random variable according to the central limit theorem (CLT), and
\begin{eqnarray*}
\mathbb{E}[x[\ell]x^*[\ell']] \hspace{-0.5cm}&&= \mathbb{E}\left[\frac{1}{M}\sum_{k=0}^{M-1}s[k]\exp\left(j\frac{2\pi k \ell}{FM}\right)\frac{1}{M}\sum_{k'=0}^{M-1}s^*[k']\exp\left(-j\frac{2\pi k' \ell'}{FM}\right)\right] \\
\hspace{-0.5cm}&&= \frac{\sigma^2_s}{M^2}\sum_{k=0}^{M-1}\exp\left(j\frac{2\pi k (\ell-\ell')}{FM}\right).
\end{eqnarray*}
where the second equality follows directly from \eqref{EquDvars}. We have assumed the average received power of Wi-Fi signal is $\rho_x$ in \eqref{EquDr}, it follows that
\begin{eqnarray}\label{eq:vars}
\mathbb{E}[\sqrt{\rho_x}x[\ell]\sqrt{\rho_x}x^*[\ell]] =\frac{\rho_x\sigma^2_s}{M} = \rho_x.
\end{eqnarray}
Equivalently, we can assume $\sigma^2_s=M$ to ensure that the average power of the received Wi-Fi signal is $\rho_x$.

Substituting \eqref{eq:x150} into $x_c$ gives us
\begin{eqnarray}\label{eq:xc}
x[\ell] = \sum_{q=0}^{Q-1}c_p[q]\sum_{n=0}^{N-1}z^*[n]\frac{1}{M}\sum_{k=0}^{M-1}s_\phi[k]\exp\left(j\frac{2\pi k(Nq+n)}{FM}\right),
\end{eqnarray}
Note that a subscript $\phi=\myfloor{(Nq+n)/FM}$ is added on the QAM-modulated symbol $s[k]$. We use this subscript to denote the OFDM symbol that $s[k]$ comes from.
Specifically, $x_c$ accumulates the power of QN samples. These QN samples involves $\myceil{QN/FM}$ OFDM symbols since an OFDM symbol has only FM samples. Thus, the $(Nq+n)$-th sample comes from the $\myfloor{(Nq+n)/FM}$-th OFDM symbol.

The reason why we distinguish the QAM symbols that come from different OFDM symbols is that they are independent. We can then rewrite \eqref{EquDvars} as
\begin{eqnarray}\label{EquDvars2}
\mathbb{E}[s_\phi[k]s^*_{\phi'}[k']] = \sigma^2_s\delta(k-k')\delta(\phi-\phi'),
\end{eqnarray}

The duration of one OFDM symbol corresponds to the duration of $\myceil{FM/N}$ ZC sequences, we can rewrite \eqref{eq:xc} as
\begin{eqnarray}\label{eq:xc2}
x[\ell] = \sum_{\phi=0}^{\myceil{\frac{QN}{FM}}-1}\sum_{i=0}^{\myceil{\frac{FM}{N}}-1}c_p\left[\phi\myceil{\frac{FM}{N}}+i\right]
\sum_{n=0}^{N-1}z^*[n]\frac{1}{M}\sum_{k=0}^{M-1}s_\phi[k]\exp\left(j\frac{2\pi k(Ni+n)}{FM}\right),
\end{eqnarray}
and
\begin{eqnarray}\label{eq:varx}
&&\mathbb{E}[x_c x^*_c]= \mathbb{E}\Bigg[
\sum_{\phi=0}^{\myceil{\frac{QN}{FM}}-1}\sum_{i=0}^{\myceil{\frac{FM}{N}}-1}c_p\left[\phi\myceil{\frac{FM}{N}}+i\right]
\sum_{n=0}^{N-1}z^*[n]\frac{1}{M}\sum_{k=0}^{M-1}s_\phi[k]\exp\left(j\frac{2\pi k(Ni+n)}{FM}\right) \nonumber\\
&&\times\sum_{\phi'=0}^{\myceil{\frac{QN}{FM}}-1}\sum_{i'=0}^{\myceil{\frac{FM}{N}}-1}c_p\left[\phi'\myceil{\frac{FM}{N}}+i'\right]
\sum_{n'=0}^{N-1}z[n']\frac{1}{M}\sum_{k'=0}^{M-1}s^*_{\phi'}[k']\exp\left(-j\frac{2\pi k'(Ni'+n')}{FM}\right)
\Bigg] \nonumber \\
&&=\frac{\sigma^2_s}{M^2}\sum_{\phi=0}^{\myceil{\frac{QN}{FM}}-1}\sum_{i=0}^{\myceil{\frac{FM}{N}}-1}c_p\left[\phi\myceil{\frac{FM}{N}}+i\right]
\sum_{i'=0}^{\myceil{\frac{FM}{N}}-1}c_p\left[\phi'\myceil{\frac{FM}{N}}+i'\right]\nonumber\\
&&\hspace{5cm} \times\sum_{n=0}^{N-1}z^*[n]\sum_{n'=0}^{N-1}z[n']\sum_{k=0}^{M-1}\exp\left(j\frac{2\pi k(Ni-Ni'+n-n')}{FM}\right) \nonumber\\
&&=\frac{1}{M}\sum_{\phi=0}^{\myceil{\frac{QN}{FM}}-1}\sum_{i=0}^{\myceil{\frac{FM}{N}}-1}c_p\left[\phi\myceil{\frac{FM}{N}}+i\right]
\sum_{i'=0}^{\myceil{\frac{FM}{N}}-1}c_p\left[\phi'\myceil{\frac{FM}{N}}+i'\right]\nonumber\\
&&\hspace{5cm} \times\sum_{n=0}^{N-1}z^*[n]\sum_{n'=0}^{N-1}z[n']\sum_{k=0}^{M-1}\exp\left(j\frac{2\pi k(Ni-Ni'+n-n')}{FM}\right) \nonumber\\
&&\triangleq \sigma^2_x,
\end{eqnarray}
where the second equality follows from \eqref{EquDvars2}, and the third equality follows from \eqref{eq:vars}. Further simplifying \eqref{eq:varx} is non-trivial.

Numerically, computing $\sigma^2_x$ from \eqref{eq:varx} is intensive. An approximation to \eqref{eq:varx} is given by
\begin{eqnarray}\label{eq:varx2}
\sigma^2_x \hspace{-0.5cm}&&\approx \mathbb{E}\Bigg[
\sum_{q=0}^{Q-1}c_p\left[q\right] \sum_{n=0}^{N-1}z^*[n]\frac{1}{M}\sum_{k=0}^{M-1}s_q[k]\exp\left(j\frac{2\pi k(Nq+n)}{FM}\right) \nonumber\\
&&\hspace{3cm}\times \sum_{q'=0}^{Q-1}c_p\left[q'\right] \sum_{n'=0}^{N-1}z^*[n']\frac{1}{M}\sum_{k'=0}^{M-1}s_{q'}[k']\exp\left(j\frac{2\pi k'(Nq'+n')}{FM}\right)
\Bigg] \nonumber \\
\hspace{-0.5cm}&&=\frac{Q}{M} \sum_{n=0}^{N-1}z^*[n]\sum_{n'=0}^{N-1}z[n']\sum_{k=0}^{M-1}\exp\left(j\frac{2\pi k(n-n')}{FM}\right).
\end{eqnarray}
This is a conservative approximation because we have assumed $s[k]$ are independent for different $q$. As a result, the approximated $\sigma^2_x$ in \eqref{eq:varx2} is larger than the real $\sigma^2_x$ in \eqref{eq:varx}. Yet, the computational complexity is greatly reduced.

\end{NewProof}

\section{The optimal threshold and the MDR of Neyman-Pearson detector}\label{sec:appC}
This appendix proves Theorem \ref{thm:4}.

\begin{NewProof}
To derive the optimal threshold and the corresponding MDR, we have to analyze $P(u\mid H_0)$ and $P(u\mid H_1)$.

Given Lemma \ref{thm:2} and \ref{thm:3}, we have $x_c\sim\mathcal{CN}(0,\sigma_x^2)$ and $w_c\sim\mathcal{CN}(0,\sigma_w^2)$. Thus,
\begin{eqnarray*}
u_c = \sqrt{\rho_x}x_c + w_c \sim\mathcal{CN}(0,\sigma_u^2=\rho_x\sigma^2_x+NQ\sigma^2_w),
\end{eqnarray*}
because $x_c$ and $w_c$ are independent.

If $H_0$ is true, $u$ follows Rayleigh distribution, giving
\begin{eqnarray}\label{eq:u0}
u = \left|u_c\right|\sim\textup{Rayleigh}(\sigma_u),
\end{eqnarray}
and if $H_1$ is true, $u$ follows Rician distribution, giving
\begin{eqnarray}\label{eq:u1}
u = \left|\sqrt{\rho_I}NQ+u_c\right|\sim\textup{Rice}(\sigma_u),
\end{eqnarray}

From \eqref{eq:u0} and \eqref{eq:u1}, we can compute the likelihood functions of $u$, as follows:
\begin{eqnarray*}
P(u\mid H_0) \hspace{-0.5cm}&&= \frac{u}{\sigma^2_u}\exp\left(-\frac{u^2}{2\sigma^2_u} \right), \\
P(u\mid H_1) \hspace{-0.5cm}&&= \frac{u}{\sigma^2_u}\exp\left(-\frac{u^2+\rho_I N^2Q^2}{2\sigma^2_u}\right)I_0\left(\frac{\sqrt{\rho_I}NQ}{\sigma^2_u}u \right),
\end{eqnarray*}
where $I_0(*)$ is the modified Bessel function of the first kind with order zero. The likelihood ratio
\begin{eqnarray}\label{eq:lambdau}
\lambda(u)=\frac{P(u\mid H_1)}{P(u\mid H_0)}=\exp\left(-\frac{\rho_I N^2Q^2}{2\sigma^2_u}\right)I_0\left(\frac{\sqrt{\rho_I}NQ}{\sigma^2_u}u \right)
\end{eqnarray}

Substituting \eqref{eq:lambdau} into \eqref{eq:rule1} gives us
\begin{eqnarray}\label{eq:rule2}
I_0\left(\frac{\sqrt{\rho_I}NQ}{\sigma^2_u}u \right)
\mathop{\lessgtr}_{H_1}^{H_0}
\exp\left(\frac{\rho_I N^2Q^2}{2\sigma^2_u}\right)\gamma
\triangleq \Gamma
\end{eqnarray}

\begin{figure}[t]
  \centering
  \includegraphics[width=0.4\columnwidth]{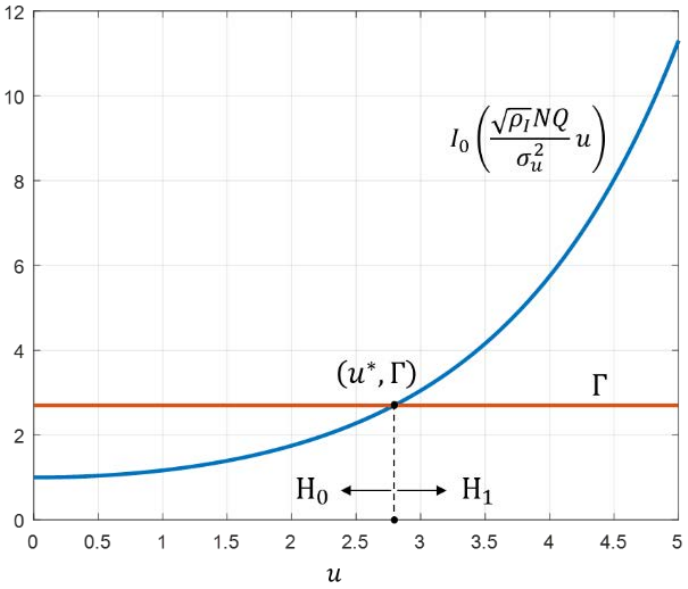}\\
  \caption{Illustration of the decision rule of Neyman-Pearson detector. $\Gamma$ is a threshold to the likelihood ratio, while $u^*$ is the corresponding threshold to the observation $u$.}
\label{fig:rule}
\end{figure}

The new decision rule in \eqref{eq:rule2} is illustrated in Fig.~\ref{fig:rule}. Given an observation $u$, we compute the modified Bessel function on the LHS of \eqref{eq:rule2}, and compare it with a threshold $\Gamma$ independent of the observation $u$. In particular, the $\Gamma$ is chosen such that the FAR $P_F=\alpha$. As shown in Fig.~\ref{fig:rule}, let the intersection of the two curves be $(u^*,\Gamma)$, thus we have
\begin{eqnarray*}
P_F=\int_{u^*}^{\infty}P(u\mid H_0)du=\int_{u^*}^{\infty} \frac{u}{\sigma^2_u}\exp\left(-\frac{u^2}{2\sigma^2_u} \right)du=\exp\left(-\frac{(u^*)^2}{2\sigma^2_u} \right)=\alpha.
\end{eqnarray*}
This gives us
\begin{eqnarray}\label{eq:ustar}
u^* = \sqrt{-2\sigma^2_u\ln \alpha}
\end{eqnarray}

We can then compute the MDR by
\begin{eqnarray}\label{eq:PM}
P_M\hspace{-0.5cm}&&=\int_{0}^{u^*}P(u\mid H_1)du=\int_{0}^{u^*} \exp\left(-\frac{u^2+\rho_I N^2Q^2}{2\sigma^2_u}\right)I_0\left(\frac{\sqrt{\rho_I}NQ}{\sigma^2_u}u \right) du \nonumber\\
\hspace{-0.5cm}&&=1 - Q_1\left(\frac{\sqrt{\rho_I} NQ}{\sigma_u},\frac{u^*}{\sigma_u} \right),
\end{eqnarray}
where $Q_1$ is a Marcum Q-function.

Given a tolerable FAR $P_F=\alpha$, the optimal $u^*$ is given by \eqref{eq:ustar}, and we can compute the MDR numerically as per \eqref{eq:PM}.

\end{NewProof}

\section{Message loss rate of multi-replica ALOHA}\label{sec:appD}
This appendix proves Theorem \ref{thm:5}.

\begin{NewProof}
To derive the message loss rate, we first analyze the probability $P_0$.
For any two nodes $A$ and $B$, $P_0$ is the probability that one of $A$'s replicas, say $A_1$, does not collide with $B$'s $d$ replicas $\{B_1,\allowbreak B_2,\allowbreak\cdots,\allowbreak B_d\}$. We will show that $P_0$ is essentially the proportion of the volume of an irregular polyhedron to that of a regular polyhedron.

Let us consider a simple case where $d=1$, and $P_0$ is the probability that $A_1$ does not collide with $B_1$.
Denote by $t(A_1)$ the transmission start time of packet $A_1$, and $t(B_1)$ the transmission start time of packet $B_1$. To avoid collisions, $t(A_1)$ and $t(B_1)$ must satisfy the following constraints:
\begin{eqnarray}\label{eq:appB1}
\left\{
\begin{array}{l}
\left| t(A_1)-t(B_1)\right| \geq T_p, \\
0\leq t(A_1)\leq T-T_p,  ~
0\leq t(B_1)\leq T-T_p.
\end{array} \right. \nonumber
\end{eqnarray}

\begin{figure}[t]
  \centering
  \includegraphics[width=0.7\columnwidth]{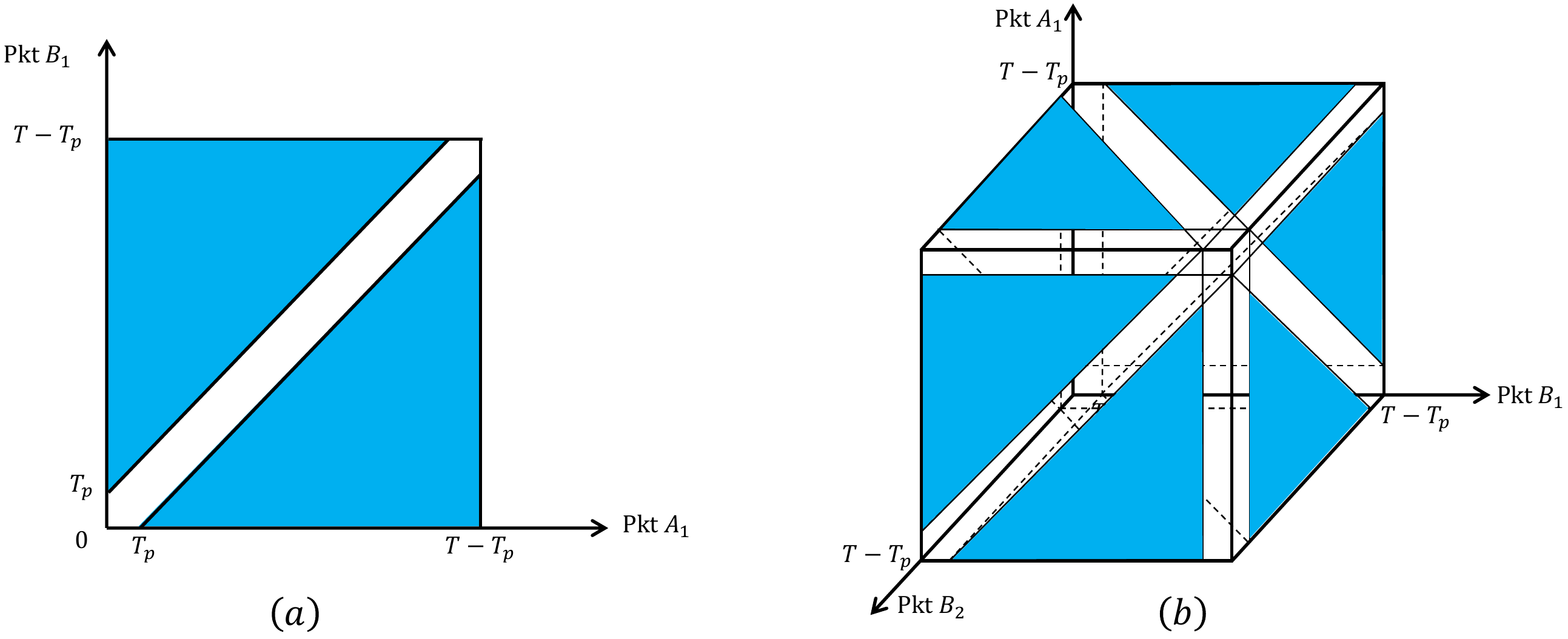}\\
  \caption{Illustration of $P_0$ when $d=1$ and $d = 2$.}
\label{Fig11}
\end{figure}

Fig.~\ref{Fig11} (a) illustrates these constraints, wherein the shaded regions are the regions that satisfy the constraints. $P_0$ is then the proportion of the shaded area to the total area, giving,
\begin{eqnarray}
P_0=\frac{(T-2T_p)^2}{(T-T_p)^2}.\nonumber
\end{eqnarray}

Next, we consider the case $d=2$, and $P_0$ is the probability that $A_1$ does not collide with $B_1$ and $B_2$. To avoid collisions, the transmission start times $t(A_1)$, $t(B_1)$ and $t(B_2)$ must satisfy
\begin{eqnarray}
\left\{
\begin{array}{l}
\left| t(A_1)-t(B_1)\right| \geq T_p,~~
\left| t(A_1)-t(B_2)\right| \geq T_p, ~~
\left| t(B_1)-t(B_2)\right| \geq T_p, \\
0\leq t(A_1)\leq T-T_p,   ~~
0\leq t(B_1)\leq T-T_p,  ~~
0\leq t(B_2)\leq T-T_p.
\end{array} \right.\nonumber
\end{eqnarray}
In particular, the condition $\left| t(B_1)-t(B_2)\right| \geq T_p$ is met by default because node $B$ will not transmit two overlapping packets.
$P_0$ can be derived as
\begin{eqnarray}\label{eq:appB3}
P_0 = \textup{Pr}(A_1,B_1,B_2 \textup{~do~not~collide} \mid B_1,B_2 \textup{~do~not~collide})
=\frac{\textup{Pr}(A_1,B_1,B_2 \textup{~do~not~collide})}{\textup{Pr}(B_1,B_2 \textup{~do~not~collide})}.
\end{eqnarray}

Fig~\ref{Fig11} (b) illustrates the regions associated with the numerator and denominator of \eqref{eq:appB3}. As can be seen, the region associated with the numerator of \eqref{eq:appB3} is essentially a cube with side length $T-3T_p$. On the other hand, the region associated with the denominator of \eqref{eq:appB3} is a cuboid with length $T-2T_p$ ($x$-axis), width $T-2T_p$ ($y$-axis) and height $T-T_p$ ($z$-axis). As a result,
\begin{eqnarray}
P_0=\frac{(T-3T_p)^3}{(T-2T_p)^2(T-T_p)}.\nonumber
\end{eqnarray}

In general, for the general case where node $B$ transmit $d$ replicas, we have
\begin{eqnarray}
P_0 \!\!\!\!\!&=&\!\!\!\!\! \textup{Pr}(\!A_1,\!B_1,...,\!B_d \textup{\!~do\!~not\!~collide} \!\mid\! B_1,...,\!B_d \textup{\!~do\!~not\!~collide}) \nonumber\\
&=&\!\!\!\!\!\frac{\textup{Pr}(A_1,B_1,...,B_d \textup{~do~not~collide})}{\textup{Pr}(B_1,...,B_d \textup{~do~not~collide})}.\nonumber
\end{eqnarray}
where the numerator represents a ($d+1$)-dimensional regular polyhedron with side length $T-(d+1)T_p$, and the denominator represents a  ($d+1$)-dimensional polyhedron with side length $T-dT_p$, $T-dT_p$, ..., $T-dT_p$, $T-T_p$ (i.e., only one side is of length $T-T_p$).
Thus, we have
\begin{eqnarray}
P_0=\frac{[T-(d+1)T_p]^{d+1}}{(T-dT_p)^d(T-T_p)}.\nonumber
\end{eqnarray}

$P_0$ is the probability that $A_1$ is clean with respect to $B$'s messages. By the assumption that each node broadcasts the packet independently, $P_0$ is also the probability that $A_1$ is clean with respect to any other node's messages and that $P_0^{K-1}$ is the probability that $A_1$ is clean with respect to all other node's messages.

The probability of $A_1$ experiencing a collision is therefore $1-P_0^{K-1}$. If we make an approximating assumption that the collision events of the $d$ replicas of node $A$ are independent, then we can write
\begin{eqnarray}
R_{\textup{loss}}\approx(1-P_0^{K-1})^d.
\end{eqnarray}
where $R_{\textup{loss}}$ is the probability that all $d$ replicas of node $A$ fail to be transmitted successfully. That is, $R_{\textup{loss}}$ is the message loss rate.

\end{NewProof}

\section{Reconciling \eqref{eq:R_loss} with \eqref{eq:R_loss_new}}\label{sec:appF}
To reconcile the $R_\textup{loss}$ derived in \eqref{eq:R_loss} and \eqref{eq:R_loss_new}, we want to show that, for the regime of our interest (i.e., $T\gg dT_p$, and $K\gg 1$), $P^{K-1}_0$ in \eqref{eq:R_loss} is approximately equal to $e^{-\rho d}$ given in \eqref{eq:R_loss_new}, where $\rho=\frac{2(K-1)T_p}{T}$.

From \eqref{eq:P0}, we have
\begin{eqnarray}
P^{K-1}_0\!\!\!&=&\!\!\!\left(\frac{1-\frac{(d+1)T_p}{T}}{1-\frac{dT_p}{T}} \right)^{d(K-1)}
\left(\frac{1-\frac{(d+1)T_p}{T}}{1-\frac{T_p}{T}}\right)^{K-1}
=\left(1-\frac{\frac{T_p}{T}}{1-\frac{dT_p}{T}} \right)^{d(K-1)}
\left(1-\frac{\frac{dT_p}{T}}{1-\frac{T_p}{T}} \right)^{K-1} \nonumber\\
&\approx&\!\!\!\left(1-\frac{T_p}{T}\right)^{d(K-1)}\left(1-\frac{dT_p}{T}\right)^{K-1}. \nonumber
\end{eqnarray}
where the approximation follows because $\frac{dT_p}{T}\ll 1$.

As $K \rightarrow~\infty$, we have
\begin{eqnarray}
\lim_{K\rightarrow\infty}P^{K-1}_0 = e^{-\frac{d(K-1)T_p}{T}}e^{-\frac{d(K-1)T_p}{T}} = e^{-\frac{2d(K-1)T_p}{T}}. \nonumber
\end{eqnarray}

\end{document}